\documentclass[twocolumn, twocolappendix]{aastex702}

\newcommand{\eg}{\textit{e.g.}}
\newcommand{\ie}{\textit{i.e.}}
\newcommand{\vs}{\textit{vs.}}
\newcommand{\oii}{[\ion{O}{2}]}
\newcommand{\oiilam}{[\ion{O}{2}]~\ensuremath{\lambda\lambda3726,29}}

\newcommand{\hbeta}{H\ensuremath{\beta}}
\newcommand{\mgii}{\ion{Mg}{2}}

\begin{document}

\title{
Redrock: Spectroscopic Classification and Redshift Fitting\\for the Dark Energy Spectroscopic Instrument
}

\author[0000-0003-4162-6619]{S. Bailey}
\affiliation{Lawrence Berkeley National Laboratory, 1 Cyclotron Road, Berkeley, CA 94720, USA}
\email{stephenbailey@lbl.gov}

\author[0000-0002-8934-0954]{A.~Brodzeller}
\affiliation{Lawrence Berkeley National Laboratory, 1 Cyclotron Road, Berkeley, CA 94720, USA}
\email{AllysonBrodzeller@lbl.gov}

\author[0000-0002-2733-4559]{J.~Moustakas}
\affiliation{Department of Physics and Astronomy, Siena University, 515 Loudon Road, Loudonville, NY 12211, USA}
\email{jmoustakas@siena.edu}

\author[0000-0003-2923-1585]{A.~Anand}
\affiliation{Inter-University Centre for Astronomy and Astrophysics, Post Bag 4, Ganeshkhind, Pune 411 007, India}
\affiliation{Lawrence Berkeley National Laboratory, 1 Cyclotron Road, Berkeley, CA 94720, USA}
\email{abhijeetanand2011@gmail.com}

\author{C. Warner}
\affiliation{Department of Astronomy, University of Florida, 211 Bryant Space Science Center, Gainesville, FL 32611, USA}
\email{cwarner@astro.ufl.edu}

\author[0000-0003-3510-7134]{T.~Kisner}
\affiliation{Lawrence Berkeley National Laboratory, 1 Cyclotron Road, Berkeley, CA 94720, USA}
\email{tskisner@lbl.gov}


\author{J.~Aguilar}
\affiliation{Lawrence Berkeley National Laboratory, 1 Cyclotron Road, Berkeley, CA 94720, USA}
\email{jaguilar@lbl.gov}

\author[0000-0001-6098-7247]{S.~Ahlen}
\affiliation{Department of Physics, Boston University, 590 Commonwealth Avenue, Boston, MA 02215 USA}
\email{ahlen@bu.edu}

\author[0000-0002-5896-6313]{D.~M.~Alexander}
\affiliation{Centre for Extragalactic Astronomy, Department of Physics, Durham University, South Road, Durham, DH1 3LE, UK}
\email{d.m.alexander@durham.ac.uk}

\author[0000-0002-0084-572X]{C.~Allende~Prieto}
\affiliation{Departamento de Astrof\'{\i}sica, Universidad de La Laguna (ULL), E-38206, La Laguna, Tenerife, Spain}
\email{carlos.allende.prieto@iac.es}

\author[0000-0001-7600-5148]{E.~Armengaud}
\affiliation{IRFU, CEA, Universit\'{e} Paris-Saclay, F-91191 Gif-sur-Yvette, France}
\email{eric.armengaud@cea.fr}

\author[0000-0001-5998-3986]{A.~Aviles}
\affiliation{Instituto Avanzado de Cosmolog\'{\i}a A.~C., San Marcos 11 - Atenas 202. Magdalena Contreras. Ciudad de M\'{e}xico C.~P.~10720, M\'{e}xico}
\affiliation{Instituto de Ciencias F\'{\i}sicas, Universidad Nacional Aut\'onoma de M\'exico, Av. Universidad s/n, Cuernavaca, Morelos, C.~P.~62210, M\'exico}
\email{aviles@icf.unam.mx}

\author[0000-0003-0467-5438]{F.~Beutler}
\affiliation{Institute for Astronomy, University of Edinburgh, Royal Observatory, Blackford Hill, Edinburgh EH9 3HJ, UK}
\email{florian.beutler@ed.ac.uk}

\author[0000-0001-9712-0006]{D.~Bianchi}
\affiliation{Dipartimento di Fisica ``Aldo Pontremoli'', Universit\`a degli Studi di Milano, Via Celoria 16, I-20133 Milano, Italy}
\affiliation{INAF-Osservatorio Astronomico di Brera, Via Brera 28, 20122 Milano, Italy}
\email{davide.bianchi1@unimi.it}

\author[0000-0002-9836-603X]{A.~S.~Bolton}
\affiliation{SLAC National Accelerator Laboratory, 2575 Sand Hill Road, Menlo Park, CA 94025, USA}
\email{abolton@slac.stanford.edu}

\author{D.~Brooks}
\affiliation{Department of Physics \& Astronomy, University College London, Gower Street, London, WC1E 6BT, UK}
\email{david.brooks@ucl.ac.uk}

\author[0000-0003-3044-5150]{A.~Carnero Rosell}
\affiliation{Departamento de Astrof\'{\i}sica, Universidad de La Laguna (ULL), E-38206, La Laguna, Tenerife, Spain}
\affiliation{Instituto de Astrof\'{\i}sica de Canarias, C/ V\'{\i}a L\'{a}ctea, s/n, E-38205 La Laguna, Tenerife, Spain}
\email{acarnero@iac.es}

\author[0000-0001-7316-4573]{F.~J.~Castander}
\affiliation{Institut d'Estudis Espacials de Catalunya (IEEC), c/ Esteve Terradas 1, Edifici RDIT, Campus PMT-UPC, 08860 Castelldefels, Spain}
\affiliation{Institute of Space Sciences, ICE-CSIC, Campus UAB, Carrer de Can Magrans s/n, 08913 Bellaterra, Barcelona, Spain}
\email{fjc@ice.csic.es}

\author[0000-0001-8996-4874]{E.~Chaussidon}
\affiliation{Lawrence Berkeley National Laboratory, 1 Cyclotron Road, Berkeley, CA 94720, USA}
\email{echaussidon@lbl.gov}

\author{T.~Claybaugh}
\affiliation{Lawrence Berkeley National Laboratory, 1 Cyclotron Road, Berkeley, CA 94720, USA}
\email{tmclaybaugh@lbl.gov}

\author[0000-0002-4213-8783]{Tamara~M.~Davis}
\affiliation{School of Mathematics and Physics, University of Queensland, Brisbane, QLD 4072, Australia}
\email{tamarad@physics.uq.edu.au}

\author[0000-0002-0553-3805]{K.~S.~Dawson}
\affiliation{Department of Physics and Astronomy, The University of Utah, 115 South 1400 East, Salt Lake City, UT 84112, USA}
\email{kdawson@astro.utah.edu}

\author[0000-0002-1769-1640]{A.~de la Macorra}
\affiliation{Instituto de F\'{\i}sica, Universidad Nacional Aut\'{o}noma de M\'{e}xico,  Circuito de la Investigaci\'{o}n Cient\'{\i}fica, Ciudad Universitaria, Cd. de M\'{e}xico  C.~P.~04510,  M\'{e}xico}
\email{macorra@fisica.unam.mx}

\author[0000-0002-5665-7912]{Biprateep~Dey}
\affiliation{Department of Astronomy \& Astrophysics, University of Toronto, Toronto, ON M5S 3H4, Canada}
\affiliation{Department of Physics \& Astronomy and Pittsburgh Particle Physics, Astrophysics, and Cosmology Center (PITT PACC), University of Pittsburgh, 3941 O'Hara Street, Pittsburgh, PA 15260, USA}
\email{b.dey@utoronto.ca}

\author[0000-0003-1251-532X]{V.~A.~Fawcett}
\affiliation{European Southern Observatory, Karl-Schwarzschild-Stra{\ss}e 2, 85748 Garching bei M\"unchen, Germany}
\email{victoria.fawcett@eso.org}

\author[0000-0003-4992-7854]{S.~Ferraro}
\affiliation{Lawrence Berkeley National Laboratory, 1 Cyclotron Road, Berkeley, CA 94720, USA}
\affiliation{University of California, Berkeley, 110 Sproul Hall \#5800 Berkeley, CA 94720, USA}
\email{sferraro@lbl.gov}

\author[0000-0002-3033-7312]{A.~Font-Ribera}
\affiliation{Instituci\'{o} Catalana de Recerca i Estudis Avan\c{c}ats, Passeig de Llu\'{\i}s Companys, 23, 08010 Barcelona, Spain}
\affiliation{Institut de F\'{i}sica d’Altes Energies (IFAE), The Barcelona Institute of Science and Technology, Edifici Cn, Campus UAB, 08193 Bellaterra (Barcelona), Spain}
\email{afont@ifae.es}

\author[0000-0002-2890-3725]{J.~E.~Forero-Romero}
\affiliation{Departamento de F\'isica, Universidad de los Andes, Cra. 1 No. 18A-10, Edificio Ip, CP 111711, Bogot\'a, Colombia}
\email{je.forero@uniandes.edu.co}

\author[0000-0001-9632-0815]{E.~Gaztañaga}
\affiliation{Institut d'Estudis Espacials de Catalunya (IEEC), c/ Esteve Terradas 1, Edifici RDIT, Campus PMT-UPC, 08860 Castelldefels, Spain}
\affiliation{Institute of Cosmology and Gravitation, University of Portsmouth, Dennis Sciama Building, Portsmouth, PO1 3FX, UK}
\affiliation{Institute of Space Sciences, ICE-CSIC, Campus UAB, Carrer de Can Magrans s/n, 08913 Bellaterra, Barcelona, Spain}
\email{gaztanaga@gmail.com}

\author[0000-0003-3142-233X]{Satya~{Gontcho A Gontcho}}
\affiliation{University of Virginia, Department of Astronomy, Charlottesville, VA 22904, USA}
\email{satya@virginia.edu}

\author[0000-0002-0676-3661]{D.~Green}
\affiliation{Lawrence Berkeley National Laboratory, 1 Cyclotron Road, Berkeley, CA 94720, USA}
\email{dylangreen@lbl.gov}

\author{G.~Gutierrez}
\affiliation{Fermi National Accelerator Laboratory, PO Box 500, Batavia, IL 60510, USA}
\email{gaston@fnal.gov}

\author[0000-0001-9822-6793]{J.~Guy}
\affiliation{Lawrence Berkeley National Laboratory, 1 Cyclotron Road, Berkeley, CA 94720, USA}
\email{jguy@lbl.gov}

\author[0000-0002-2312-3121]{B.~Hadzhiyska}
\affiliation{Institute of Astronomy, University of Cambridge, Madingley Road, Cambridge CB3 0HA, UK}
\affiliation{University of California, Berkeley, 110 Sproul Hall \#5800 Berkeley, CA 94720, USA}
\email{boryanah@alumni.princeton.edu}

\author[0000-0003-1197-0902]{C.~Hahn}
\affiliation{Department of Astronomy, University of Texas at Austin, 2515 Speedway, TX 78712, USA}
\email{changhoon.hahn@utexas.edu}

\author[0000-0002-9136-9609]{H.~K.~Herrera-Alcantar}
\affiliation{Institut d'Astrophysique de Paris. 98 bis boulevard Arago. 75014 Paris, France}
\affiliation{Center for Cosmology and AstroParticle Physics, The Ohio State University, 191 West Woodruff Avenue, Columbus, OH 43210, USA}
\affiliation{Department of Astronomy, The Ohio State University, 4055 McPherson Laboratory, 140 W 18th Avenue, Columbus, OH 43210, USA}
\affiliation{IRFU, CEA, Universit\'{e} Paris-Saclay, F-91191 Gif-sur-Yvette, France}
\affiliation{The Ohio State University, Columbus, 43210 OH, USA}
\email{herrera.301@osu.edu}

\author[0000-0002-6550-2023]{K.~Honscheid}
\affiliation{Center for Cosmology and AstroParticle Physics, The Ohio State University, 191 West Woodruff Avenue, Columbus, OH 43210, USA}
\affiliation{Department of Physics, The Ohio State University, 191 West Woodruff Avenue, Columbus, OH 43210, USA}
\affiliation{The Ohio State University, Columbus, 43210 OH, USA}
\email{kh@physics.osu.edu}

\author[0000-0001-6558-0112]{D.~Huterer}
\affiliation{Department of Physics, University of Michigan, 450 Church Street, Ann Arbor, MI 48109, USA}
\affiliation{University of Michigan, 500 S. State Street, Ann Arbor, MI 48109, USA}
\email{huterer@umich.edu}

\author[0000-0002-6024-466X]{M.~Ishak}
\affiliation{Department of Physics, The University of Texas at Dallas, 800 W. Campbell Rd., Richardson, TX 75080, USA}
\email{mishak@utdallas.edu}

\author[0000-0003-4176-6486]{L.~Jiang}
\affiliation{Kavli Institute for Astronomy and Astrophysics at Peking University, PKU, 5 Yiheyuan Road, Haidian District, Beijing 100871, P.R. China}
\email{jiangkiaa@pku.edu.cn}

\author[0000-0002-0000-2394]{S.~Juneau}
\affiliation{NSF NOIRLab, 950 N. Cherry Ave., Tucson, AZ 85719, USA}
\email{stephanie.juneau@noirlab.edu}

\author[0000-0002-8828-5463]{D.~Kirkby}
\affiliation{Department of Physics and Astronomy, University of California, Irvine, 92697, USA}
\email{dkirkby@uci.edu}

\author[0000-0001-6356-7424]{A.~Kremin}
\affiliation{Lawrence Berkeley National Laboratory, 1 Cyclotron Road, Berkeley, CA 94720, USA}
\email{akremin@lbl.gov}

\author[0000-0002-1134-9035]{O.~Lahav}
\affiliation{Department of Physics \& Astronomy, University College London, Gower Street, London, WC1E 6BT, UK}
\email{o.lahav@ucl.ac.uk}

\author[0000-0002-6731-9329]{C.~Lamman}
\affiliation{The Ohio State University, Columbus, 43210 OH, USA}
\email{lamman.1@osu.edu}

\author[0000-0001-8857-7020]{T.-W.~Lan}
\affiliation{Graduate Institute of Astrophysics and Department of Physics, National Taiwan University, No. 1, Sec. 4, Roosevelt Rd., Taipei 10617, Taiwan}
\email{twlan@ntu.edu.tw}

\author[0000-0003-1838-8528]{M.~Landriau}
\affiliation{Lawrence Berkeley National Laboratory, 1 Cyclotron Road, Berkeley, CA 94720, USA}
\email{mlandriau@lbl.gov}

\author[0000-0001-7178-8868]{L.~Le~Guillou}
\affiliation{Sorbonne Universit\'{e}, CNRS/IN2P3, Laboratoire de Physique Nucl\'{e}aire et de Hautes Energies (LPNHE), FR-75005 Paris, France}
\email{llg@lpnhe.in2p3.fr}

\author[0000-0002-3677-3617]{A.~Leauthaud}
\affiliation{Department of Astronomy and Astrophysics, UCO/Lick Observatory, University of California, 1156 High Street, Santa Cruz, CA 95064, USA}
\affiliation{Department of Astronomy and Astrophysics, University of California, Santa Cruz, 1156 High Street, Santa Cruz, CA 95065, USA}
\email{alexie@ucsc.edu}

\author[0000-0003-1887-1018]{M.~E.~Levi}
\affiliation{Lawrence Berkeley National Laboratory, 1 Cyclotron Road, Berkeley, CA 94720, USA}
\email{melevi@lbl.gov}

\author[0000-0003-4962-8934]{M.~Manera}
\affiliation{Departament de F\'{i}sica, Serra H\'{u}nter, Universitat Aut\`{o}noma de Barcelona, 08193 Bellaterra (Barcelona), Spain}
\affiliation{Institut de F\'{i}sica d’Altes Energies (IFAE), The Barcelona Institute of Science and Technology, Edifici Cn, Campus UAB, 08193 Bellaterra (Barcelona), Spain}
\email{mmanera@ifae.es}

\author[0000-0002-4279-4182]{P.~Martini}
\affiliation{Center for Cosmology and AstroParticle Physics, The Ohio State University, 191 West Woodruff Avenue, Columbus, OH 43210, USA}
\affiliation{Department of Astronomy, The Ohio State University, 4055 McPherson Laboratory, 140 W 18th Avenue, Columbus, OH 43210, USA}
\affiliation{The Ohio State University, Columbus, 43210 OH, USA}
\email{martini.10@osu.edu}

\author[0000-0002-1125-7384]{A.~Meisner}
\affiliation{NSF NOIRLab, 950 N. Cherry Ave., Tucson, AZ 85719, USA}
\email{aaron.meisner@noirlab.edu}

\author[0000-0002-6610-4836]{R.~Miquel}
\affiliation{Instituci\'{o} Catalana de Recerca i Estudis Avan\c{c}ats, Passeig de Llu\'{\i}s Companys, 23, 08010 Barcelona, Spain}
\affiliation{Institut de F\'{i}sica d’Altes Energies (IFAE), The Barcelona Institute of Science and Technology, Edifici Cn, Campus UAB, 08193 Bellaterra (Barcelona), Spain}
\email{rmiquel@ifae.es}

\author{A.~D.~Myers}
\affiliation{Department of Physics \& Astronomy, University  of Wyoming, 1000 E. University, Dept.~3905, Laramie, WY 82071, USA}
\email{amyers14@uwyo.edu}

\author[0000-0001-9070-3102]{S.~Nadathur}
\affiliation{Institute of Cosmology and Gravitation, University of Portsmouth, Dennis Sciama Building, Portsmouth, PO1 3FX, UK}
\email{seshadri.nadathur@port.ac.uk}

\author[0000-0002-1544-8946]{G.~Niz}
\affiliation{Departamento de F\'{\i}sica, DCI-Campus Le\'{o}n, Universidad de Guanajuato, Loma del Bosque 103, Le\'{o}n, Guanajuato C.~P.~37150, M\'{e}xico}
\affiliation{Instituto Avanzado de Cosmolog\'{\i}a A.~C., San Marcos 11 - Atenas 202. Magdalena Contreras. Ciudad de M\'{e}xico C.~P.~10720, M\'{e}xico}
\email{g.niz@ugto.mx}

\author[0000-0003-3188-784X]{N.~Palanque-Delabrouille}
\affiliation{IRFU, CEA, Universit\'{e} Paris-Saclay, F-91191 Gif-sur-Yvette, France}
\affiliation{Lawrence Berkeley National Laboratory, 1 Cyclotron Road, Berkeley, CA 94720, USA}
\email{npalanque-delabrouille@lbl.gov}

\author[0000-0002-0644-5727]{W.~J.~Percival}
\affiliation{Department of Physics and Astronomy, University of Waterloo, 200 University Ave W, Waterloo, ON N2L 3G1, Canada}
\affiliation{Perimeter Institute for Theoretical Physics, 31 Caroline St. North, Waterloo, ON N2L 2Y5, Canada}
\affiliation{Waterloo Centre for Astrophysics, University of Waterloo, 200 University Ave W, Waterloo, ON N2L 3G1, Canada}
\email{will.percival@uwaterloo.ca}

\author[0000-0001-7145-8674]{F.~Prada}
\affiliation{Instituto de Astrof\'{i}sica de Andaluc\'{i}a (CSIC), Glorieta de la Astronom\'{i}a, s/n, E-18008 Granada, Spain}
\email{fprada@iaa.es}

\author[0000-0002-4940-3009]{Ragadeepika~Pucha}
\affiliation{Department of Physics and Astronomy, The University of Utah, 115 South 1400 East, Salt Lake City, UT 84112, USA}
\email{dr.raga.pucha@gmail.com}

\author[0000-0001-6979-0125]{I.~P\'erez-R\`afols}
\affiliation{Departament de F\'isica, EEBE, Universitat Polit\`ecnica de Catalunya, c/Eduard Maristany 10, 08930 Barcelona, Spain}
\email{ignasi.perez.rafols@upc.edu}

\author[0000-0001-5999-7923]{A.~Raichoor}
\affiliation{Lawrence Berkeley National Laboratory, 1 Cyclotron Road, Berkeley, CA 94720, USA}
\affiliation{Universit\'e Paris Cit\'e, CNRS, Astroparticule et Cosmologie, F-75013 Paris, France}
\email{raichoor@apc.in2p3.fr}

\author[0000-0002-3500-6635]{C.~Ravoux}
\affiliation{Universit\'{e} Clermont-Auvergne, CNRS, LPCA, 63000 Clermont-Ferrand, France}
\email{corentin.ravoux@clermont.in2p3.fr}

\author[0000-0002-7522-9083]{A.~J.~Ross}
\affiliation{Center for Cosmology and AstroParticle Physics, The Ohio State University, 191 West Woodruff Avenue, Columbus, OH 43210, USA}
\affiliation{Department of Astronomy, The Ohio State University, 4055 McPherson Laboratory, 140 W 18th Avenue, Columbus, OH 43210, USA}
\affiliation{The Ohio State University, Columbus, 43210 OH, USA}
\email{ross.1333@osu.edu}

\author{G.~Rossi}
\affiliation{Department of Physics and Astronomy, Sejong University, 209 Neungdong-ro, Gwangjin-gu, Seoul 05006, Republic of Korea}
\email{graziano@sejong.ac.kr}

\author[0000-0002-0394-0896]{R.~Ruggeri}
\affiliation{Queensland University of Technology,  School of Chemistry \& Physics, George St, Brisbane 4001, Australia}
\email{rossana.ruggeri@qut.edu.au}

\author[0000-0002-1609-5687]{L.~Samushia}
\affiliation{Abastumani Astrophysical Observatory, Tbilisi, GE-0179, Georgia}
\affiliation{Department of Physics, Kansas State University, 116 Cardwell Hall, Manhattan, KS 66506, USA}
\email{lado@phys.ksu.edu}

\author[0000-0002-9646-8198]{E.~Sanchez}
\affiliation{CIEMAT, Avenida Complutense 40, E-28040 Madrid, Spain}
\email{eusebio.sanchez@ciemat.es}

\author[0000-0002-0408-5633]{C.~Saulder}
\affiliation{Max Planck Institute for Extraterrestrial Physics, Gie\ss enbachstra\ss e 1, 85748 Garching, Germany}
\email{csaulder@mpe.mpg.de}

\author[0000-0002-3569-7421]{E.~F.~Schlafly}
\affiliation{Space Telescope Science Institute, 3700 San Martin Drive, Baltimore, MD 21218, USA}
\email{eschlafly@stsci.edu}

\author{D.~Schlegel}
\affiliation{Lawrence Berkeley National Laboratory, 1 Cyclotron Road, Berkeley, CA 94720, USA}
\email{djschlegel@lbl.gov}

\author{M.~Schubnell}
\affiliation{Department of Physics, University of Michigan, 450 Church Street, Ann Arbor, MI 48109, USA}
\affiliation{University of Michigan, 500 S. State Street, Ann Arbor, MI 48109, USA}
\email{schubnel@umich.edu}

\author[0000-0003-3449-8583]{R.~Sharples}
\affiliation{Centre for Advanced Instrumentation, Department of Physics, Durham University, South Road, Durham DH1 3LE, UK}
\affiliation{Institute for Computational Cosmology, Department of Physics, Durham University, South Road, Durham DH1 3LE, UK}
\email{r.m.sharples@durham.ac.uk}

\author[0000-0002-3461-0320]{J.~Silber}
\affiliation{Lawrence Berkeley National Laboratory, 1 Cyclotron Road, Berkeley, CA 94720, USA}
\email{jhsilber@lbl.gov}

\author[0000-0002-2949-2155]{M.~Siudek}
\affiliation{Institute of Space Sciences, ICE-CSIC, Campus UAB, Carrer de Can Magrans s/n, 08913 Bellaterra, Barcelona, Spain}
\affiliation{Instituto de Astrof\'{\i}sica de Canarias, C/ V\'{\i}a L\'{a}ctea, s/n, E-38205 La Laguna, Tenerife, Spain}
\email{msiudek@iac.es}

\author[0000-0003-1704-0781]{G.~Tarl\'{e}}
\affiliation{University of Michigan, 500 S. State Street, Ann Arbor, MI 48109, USA}
\email{gtarle@umich.edu}

\author[0009-0001-2732-8431]{R.~Vaisakh}
\affiliation{Department of Physics, Southern Methodist University, 3215 Daniel Avenue, Dallas, TX 75275, USA}
\email{vvaisakh@smu.edu}

\author{B.~A.~Weaver}
\affiliation{NSF NOIRLab, 950 N. Cherry Ave., Tucson, AZ 85719, USA}
\email{benjamin.weaver@noirlab.edu}

\author[0000-0001-5146-8533]{C.~Yèche}
\affiliation{IRFU, CEA, Universit\'{e} Paris-Saclay, F-91191 Gif-sur-Yvette, France}
\email{christophe.yeche@cea.fr}

\author[0000-0001-5381-4372]{R.~Zhou}
\affiliation{Lawrence Berkeley National Laboratory, 1 Cyclotron Road, Berkeley, CA 94720, USA}
\email{rongpuzhou@lbl.gov}

\author[0000-0002-6684-3997]{H.~Zou}
\affiliation{National Astronomical Observatories, Chinese Academy of Sciences, A20 Datun Road, Chaoyang District, Beijing, 100101, P.~R.~China}
\email{zouhu@nao.cas.cn}

\begin{abstract}
Redrock is a software package for spectroscopic classification and redshift fitting, designed to use the full information from  spectroperfectionism outputs: flux-calibrated spectra, uncorrelated errors, and a per-spectrum resolution matrix encoding the non-Gaussian wavelength-dependent spectral line-spread function.
Redrock was originally developed for the Dark Energy Spectroscopic Instrument (DESI), but the Extended Baryon Oscillation Spectroscopic Survey (eBOSS) also used Redrock as their primary galaxy redshift fitter prior to DESI.
This paper presents the algorithmic details of Redrock and its performance on DESI data. On spectra from DESI DR2, Redrock achieves a redshift recovery rate of approximately 98\% for the Bright Galaxy Survey (BGS) and Luminous Red Galaxies (LRG), 95.9\% for Emission Line Galaxies (ELG), and 92.6\% for quasars (QSO), and it delivers unbiased stellar radial velocities with a precision of roughly 5 km~s$^{-1}$.
\end{abstract}


\section{Introduction}
\label{sec:intro}

The Dark Energy Spectroscopic Instrument \citep[DESI;][]{DESI2016b.Instr, DESI2016a.Science, DESI2022-Instrument, Corrector.Miller.2023, FiberSystem.Poppett.2024} is a spectroscopic
redshift survey designed to observe $\sim$60~million galaxies and quasars
over 17,000 deg$^2$ with an 8-year observing program \citep{Schlafly2023.DESISurvey}.
In addition to the sky-subtracted flux-calibrated
spectra, a key output of DESI is the spectral classification
(galaxy, quasar, star) and redshift of each observed object, included with the public data releases \citep[]{DESI2024.EDR,DESI2025.DR1}.
Redrock is a spectral classification and redshift fitting code designed
for the needs of DESI, while at the same time being open-source and generic
enough to be applicable to other spectroscopic surveys such as the
Sloan Digital Sky Survey (SDSS).

The algorithmic method in Redrock is similar to that used by
SDSS~\citep{Bolton2012}: selecting the $\chi^2$ minimum from
Principal Component Analysis (PCA) template fits over a range of redshifts
and template spectral types.  It improves upon that method through
updated templates, more rigorous error propagation, the use of
wavelength-dependent non-Gaussian per-spectrum resolution modeling, and
optional spectral archetype fits to avoid unphysical solutions \citep{Anand2024.RedrockArchetypes}.
The improvements in Redrock were significant enough that the SDSS-IV
Extended Baryon Oscillation Spectroscopic Survey (eBOSS) switched to using
Redrock for the final cosmology analyses instead of the original SDSS code
\citep{Ross2020.eBOSS}.

Other previous methods for automated spectroscopic classification and redshift measurement
include template Fourier cross-correlation \citep{TonryDavis1979},
explicitly matching observed spectral peaks to sets of known emission lines \citep[WiggleZ;][]{WiggleZ},
template fitting after subtracting a median-filtered spectrum \citep[DEEP2;][]{DEEP2},
and fitting flux-calibrated spectra to a library of archetype basis templates \citep[PRIMUS;][]{cool13-primus}.
Redrock also uses flux-calibrated spectra --- including the continuum --- and fits templates across all observed wavelengths simultaneously without explicitly searching for and fitting individual emission or absorption features.  
This approach enables Redrock to use the full information available in the data.
Since the PCA templates are derived from complete spectra, they implicitly encode correlations between line strengths, relative velocity offsets (for the quasar templates), and other line properties in addition to the relationship between these features and the underlying continuum shape. In principle, this offers an advantage over methods that rely on identification of individual features or groups of features: many low signal-to-noise emission and/or absorption lines can collectively contribute to achieving a secure classification and emission line confusion is discouraged. However, this approach is limited by the adequacy of the PCA templates, and reliant upon upstream processing to provide high-quality sky-subtracted flux-calibrated spectra with well-characterized errors as input.

Redrock comes with a set of templates for fitting stars, quasars (QSOs), and galaxies.
The default templates are tuned to the primary populations targeted by DESI. For example, the default galaxy templates are derived from simulated spectra of nearby bright galaxies \citep[BGS;][]{BGS.TS.Hahn.2023}, old metal-rich luminous red galaxies \citep[LRG;][]{LRG.TS.Zhou.2023}, and active star-forming emission line galaxies \citep[ELG;][]{ELG.TS.Raichoor.2023} while the quasar templates are designed to span the spectral diversity observed over the full targeted redshift range ($z<3.5$) for cosmology  \citep{QSO.TS.Chaussidon.2023}.
At the same time, the templates are kept separate from the code, enabling future projects to develop their own custom templates that match their target populations \citep[e.g.,][]{ruhlmann-kleider2024}.

This paper is organized as follows.
\S\ref{sec:overview} provides an overview of the Redrock algorithmic method.
\S\ref{sec:templates} describes the underlying galaxy, QSO, and stellar templates, with further details in Appendix~\ref{appendix:galtemplates}.
\S\ref{sec:algorithm_performance} lists the performance of Redrock on DESI data.
Known issues are described in \S\ref{sec:known_issues}.
\S\S \ref{sec:computational_performance}, \ref{sec:outputs}, \ref{sec:code}
discuss the computational performance, output files, and Redrock code, respectively.  We conclude in \S\ref{sec:conclusions}.

\section{Overview of Method}
\label{sec:overview}

Redrock determines spectral classes and redshifts of astronomical spectra by fitting
Principal Component Analysis (PCA) templates to the data.  Each template
set models a particular spectral class --- galaxy, quasar, or star --- and is fit over a range of redshifts $z$ to form a $\chi^2$ vs.~$z$ scan.
The minimum $\chi^2$ from these scans is selected as the best-fit spectral
classification and redshift.  In addition to their overall classification,
templates can optionally have a subtype, \eg, Redrock has multiple template
sets for different stellar subtypes; see \S\ref{sec:star-templates}.
The core Redrock algorithms are independent of the specific templates and redshift ranges used.
\S\ref{sec:templates} describes the default DESI-centric templates provided with Redrock,
but other users can develop (and have developed) their own templates to use with Redrock for other cases.

Redrock is designed to make full use of the spectral outputs of the ``spectroperfectionism'' algorithm described in \cite{Bolton2010} ---
flux-calibrated spectra with an error model that is by construction
uncorrelated across wavelength bins, and
a resolution matrix per spectrum encoding the full non-Gaussian wavelength-dependent spectral line-spread function.
In order to preserve the fidelity of the input data error model, including
its lack of covariance between wavelength bins,
Redrock always transforms the templates to match the data (redshifting, wavelength resampling, resolution convolving), rather than transforming the data to match the templates.

Redrock fits flux-calibrated spectra but does not incorporate separate imaging-based photometric fluxes or colors as part of the fit.
Although the PCA templates provide some flexibility to absorb flux-miscalibration, the templates themselves were constructed from flux-calibrated models and data without explicitly introducing broadband mis-calibration terms into the templates.
Early algorithmic comparison testing indicated that Redrock outperformed methods that first removed broadband spectral information, implying that the use of imperfectly calibrated data was better than completely removing calibration effects. However, this improvement has not been quantified on current DESI data. It should also be noted that galactic extinction is not incorporated into Redrock fits. 

DESI runs Redrock on spectra that are pre-coadded across exposures following
the algorithm described in \cite{guy23a} \S4.11, which includes the
effective resolution matrix of the resulting coadded spectra.
An equivalent algorithm is implemented independently within Redrock and is
used if the inputs are provided as uncoadded spectra; since it does not
import the DESI pipeline code, it is not guaranteed to reproduce the
DESI pipeline coadds exactly. Redrock can simultaneously fit multiple spectra for the same object, but it will coadd them only if they are already on identical wavelength grids.
For example, DESI coadds across exposures but not across spectrograph arms,
such that each target has 3 spectra for the 3 DESI spectrograph arms
B (3600--5930 \AA), R (5600--7720 \AA), and Z (7470--9800 \AA). Redrock naturally handles the camera overlap regions since the measurements
in each arm are independent.
Redrock performs a simultaneous fit to all 3 spectra per object to determine a single best-fit classification and redshift. Since the DESI spectra are in the barycentric frame, so are the resulting redshift estimates from Redrock.

The templates are assumed to be noise-free and fully span the wavelength range of the data for any possible redshift solution.
To perform the redshift scan, Redrock multiplies the wavelength grid of
the templates by $(1+z)$ then resamples to the binning of the data using
trapezoidal integration over each wavelength bin.  It then multiplies the
templates by the wavelength-dependent resolution matrix encoding the
per-spectrum instrumental resolution $R$ provided by the spectroperfectionism
extractions \citep[see \S4.5 of][]{guy23a}.
The redshifted and resampled templates $t_i$, convolved with the resolution matrix $R$, are then fit to the spectral flux $f$ with uncertainties $\sigma$ to obtain the PCA coefficients $c_i$ and best fit $\chi^2$. 

\begin{equation}
    \chi^2(z) = \sum_\lambda \left( {
        {f_\lambda - \sum_i c_i (R t_i)_\lambda} \over \sigma_\lambda
        } \right)^2
\end{equation}

\begin{table*}[t]
\centering
\begin{tabular}{l|l|l|l|l}
Template & $\lambda_{rf}$ range (\AA) & $z_\mathrm{min}$ & $z_\mathrm{max}$ & $z$~step \\
\hline
STAR     & $3000-11000$ & $-0.002$ & $0.002$ & $\Delta z = 4 \times 10^{-5}$ \\
GALAXY   & $1228-11000$ & $-0.005$ & $1.7$ & $\Delta \log(1+z) = 3 \times 10^{-4}$ \\
QSO      & $450-10399$ & $0.05$ & $6.0$ & $\Delta \log(1+z) = 5 \times 10^{-4}$ \\
QSO-LOZ  & $1329-9632$ & $0.05$ & $1.6$ & $\Delta \log(1+z) = 5 \times 10^{-4}$ \\
QSO-HIZ  & $444-4498$ & $1.4$  & $7.0$ & $\Delta \log(1+z) = 5 \times 10^{-4}$ \\
\end{tabular}
\caption{
Rest-frame wavelength coverage, redshift ranges, and redshift step sizes used by each Redrock template type.
The DESI Early Data Release (EDR) used a single QSO template set spanning
redshifts $0.05 \le z < 6.0$; Data Release 1 (DR1) and the forthcoming DR2 use two sets
of QSO templates spanning $0.05 \le z < 1.6$ and $1.4 \le z < 7.0$.
}
\label{tab:redshift_ranges}
\end{table*}

\begin{figure}[t]
\centering
\plotone{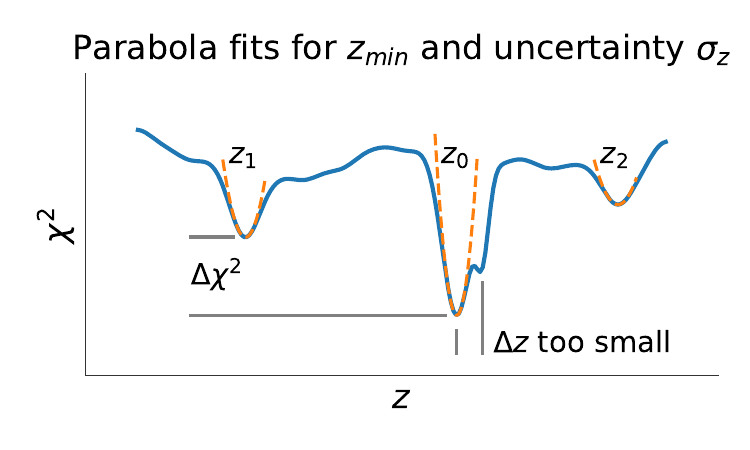}
\caption{
Schematic diagram of $\chi^2$~\vs~$z$ scan and minimum selection.
}
\label{fig:zscan_cartoon}
\end{figure}

\begin{figure}[t]
\centering
\plotone{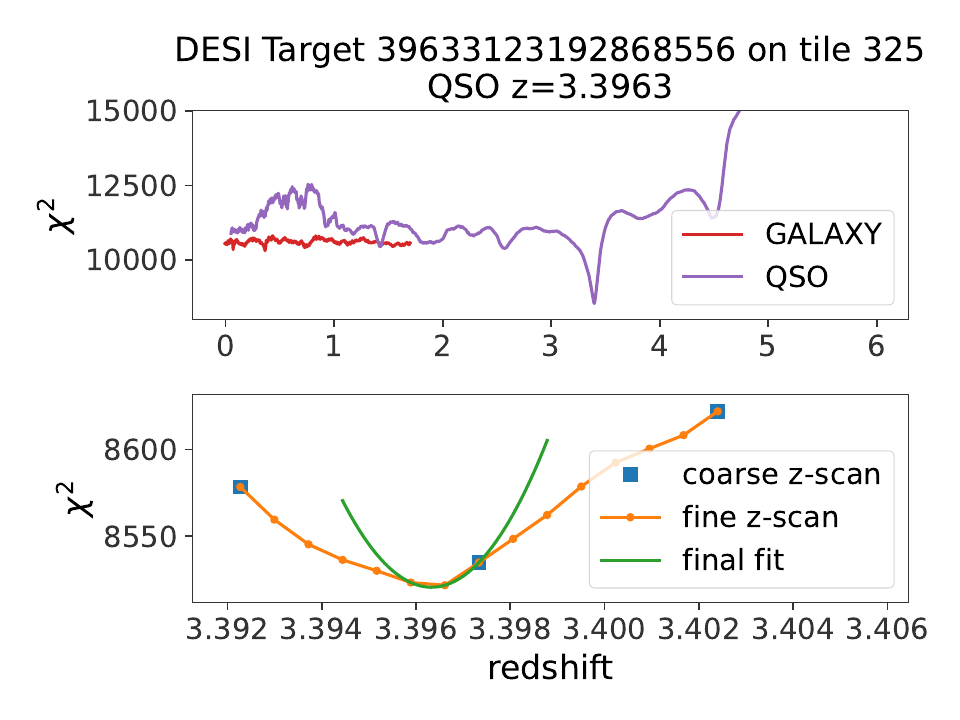}
\caption{
Example $\chi^2$~\vs~$z$ scan for a QSO.  The upper plot shows
$\chi^2$~\vs~$z$ for the
coarse redshift scan for GALAXY and QSO templates; the lower plot shows a
zoom in of the finer redshift scan around the $\chi^2$ minimum and the final
parabolic fit to the 3 points about the minimum.
}
\label{fig:zscan_minfit}
\end{figure}

The initial $\chi^2$~\vs~$z$ scan uses the redshift ranges and step sizes
given in Table~\ref{tab:redshift_ranges} for each template set.\footnote{
These ranges are valid through upcoming DESI Data Release 2; production runs for
future data releases may adjust these parameters.}
For each template set, the lowest 3 minima are rescanned using 15 finer redshift
steps spanning $\pm 1$ coarse redshift step around the original minimum.
These are refined in order from lowest $\chi^2$ local minimum to highest;
if a new minimum is within 1000~km~s$^{-1}$ of a previously fit minimum (or 100~km~s$^{-1}$ for stellar fits), it is
considered part of the same redshift solution and skipped, so that the lower-$\chi^2$ minimum is always the one retained.
A parabola is fit to the 3 points around the minimum of the new finer $\chi^2$~vs.~$z$ scan.  The parabolic fit minimum defines the
redshift and $\chi^2$ for this candidate classification/redshift,
and the change in redshift for a
$\Delta \chi^2=1$ is used as the estimate of the statistical redshift error.
This is shown schematically in Figure~\ref{fig:zscan_cartoon} for fitting
3 minima while excluding a 4th minimum that is too close in redshift to an already better minimum.
Figure~\ref{fig:zscan_minfit} shows the coarse redshift scan, refined
redshift scan, and parabola fit to the minimum for a DESI QSO target
at $z=3.3963$.
The redshift step size and number of points included in the final fit were
selected to balance algorithmic performance, robustness to fit failures,
and computational cost.

After determining the fits to the best 3 minima of each template set, these
are then compared to select the best overall redshift. Failed minima fits
are excluded, and the lowest $\chi^2$ of the remaining fits is chosen.
A fit is considered failed if its minimum is outside the range of points
being fit, the parabola has zero or negative curvature,
the $\chi^2$~\vs~$z$ parabola minimum is negative,
or the linear algebra fitting routine (\texttt{numpy.polyfit}) raises an exception (a defensive check, e.g., for non-finite $\chi^2$ values when no valid data remain).
In the case where the minimum is at an endpoint (\eg, $z=1.7$ for galaxies), the refined
scan spans only the single coarse step on the interior side, since templates
are never evaluated outside their configured redshift range. If $\chi^2$ is
still decreasing toward the boundary, the parabola minimum falls outside the
fitted points and the candidate is considered a failed fit and excluded.

Although different template sets may have different numbers of
coefficients (as is the case for the default DESI template set),
Redrock compares the total $\chi^2$, not the reduced $\chi^2$, when
selecting the best overall redshift.
While this slightly biases the solution towards templates with more eigenspectra, 
this convention was used for all DESI data productions to date.
In \S\ref{sec:known_issues} we show that ranking by reduced $\chi^2$
instead would have a negligible effect on the redshifts of confidently
classified spectra.

In addition to the single best overall classification/redshift,
Redrock also saves the original $\chi^2$~\vs~$z$ scan for each template,
and the refined scans and fits of the best 3 fits per classification
(galaxy, QSO, or star).  For example, although there are 8 different
stellar templates, only the best 3 across all stellar templates are kept,
even if the 4th best stellar fit is better than the best galaxy fit. 

\begin{figure}[t]
\centering
\plotone{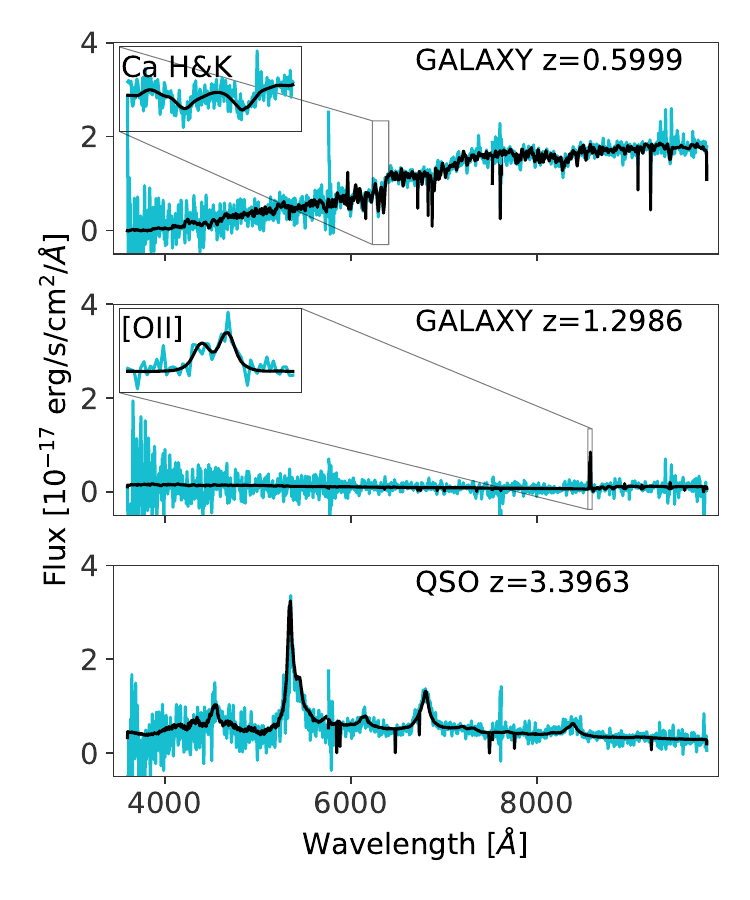}
\caption{
Example Redrock fits to an LRG (top, with inset for Ca H\&K),
an ELG (middle, with inset plot for [OII]),
and a QSO (bottom).  The targets plotted are TARGETIDs 39633118721739723 (LRG), 39633123197060417 (ELG), and 39633123192868556 (QSO)
from tile 325 of the Iron (DR1) data production. The data are shown in cyan with a 2.4~\AA~Gaussian
smoothing for visual clarity, with the best-fit template overlaid in black.  The inset plots
do not smooth the data. 
}
\label{fig:example_fits}
\end{figure}

Figure~\ref{fig:example_fits} shows Redrock fits to the 3 primary DESI
target types observed during dark time: an LRG, ELG, and QSO.  Redrock does not distinguish
between galaxy subtypes, and thus fits both the LRG and the ELG with the
same galaxy PCA templates even though they have dramatically different
continuum \vs~emission line properties. While the Redrock fits (dark lines) are to the unsmoothed data, the data in the figure (light lines) have
a 2.4~\AA~(3 wavelength bin) Gaussian smoothing for visual clarity.
Inset plots illustrate the ability of the templates to accurately model distinct galaxy features, such as the Ca H\&K absorption lines for the LRG,
and the \oiilam~doublet for the ELG.

During early Redrock development using simulated DESI data, a significant cause of incorrect redshifts was unphysical negative \oii~fits with the galaxy templates caused by noise fluctuations.
This failure case was mitigated by penalizing the $\chi^2$ for galaxy template fits if
the \oii~model flux integrated over rest-frame wavelengths
$3724 \le \lambda \le 3733$~\AA~is negative.  The penalty term is simply the
absolute value of the integrated flux in units of
$10^{-17} \mathrm{erg~s^{-1}~cm^{-2}}$.  There is no particular mathematical justification
for this choice; it pragmatically worked and development moved on.
Internal DESI studies indicate that sub-percent improvements to ELG purity could be
achieved by making this more rigorous through accounting for the measured
noise, also fitting for the continuum, and making the penalty vary with the
square of the significance instead of an unweighted linear penalty.
These possible changes have not been implemented in Redrock production code, however,
and the ad hoc linear penalty for negative \oii~remains.

\section{Default Templates}
\label{sec:templates}

Accurate spectroscopic redshifts require high-quality spectrophotometric templates which faithfully represent the spectral diversity spanned by galaxies and QSOs over cosmic time. In this section we describe the procedure used to develop the default galaxy, QSO, and stellar templates used by DESI, available
at \url{https://github.com/desihub/redrock-templates}.
Other projects could develop alternate templates for additional target types such as Lyman-$\alpha$ Emitter (LAE) galaxies.

\subsection{Principal Component Analysis Templates}
\label{sec:pca-templates}

Because Redrock always transforms the templates to match the data
(see \S\ref{sec:overview}), the templates should ideally be defined at higher
spectral resolution than the data.
As outlined in the following subsections, the galaxy and most of the stellar templates satisfy this requirement with a resolution
more than twice the maximum resolution of the DESI spectrographs and a rest-frame wavelength grid of 0.1~\AA.
The quasar templates and the cataclysmic variable
template are exceptions. They are empirically derived from SDSS/BOSS
data ($\mathcal{R}\approx1500$--2500), and thus include the
instrumental broadening of those spectra. Redrock nonetheless convolves them with the DESI resolution matrix like
any other template. For the broad emission lines that dominate quasar
redshift fits (FWHM of thousands of km~s$^{-1}$), the additional broadening
is negligible, but narrow features in the quasar templates are modeled at
somewhat lower resolution than the DESI data. This is discussed further in
\S\ref{sec:known_issues}.

\subsubsection{Galaxy templates}\label{sec:galaxy-templates}

To construct the galaxy PCA templates, we start from physically motivated galaxy spectral models. We use these models to generate a large number of synthetic galaxy spectra whose properties are matched to the observed DESI galaxy populations, and then perform PCA on this set to derive a compact set of eigenspectra spanning the range of galaxy spectra observed by DESI.

The synthetic spectra are built from the set of \textit{basis templates}
described in Appendix~\ref{appendix:galtemplates}, which span a wide range of stellar-continuum and emission-line properties. Specifically, we generate
10,000 random realizations of ELGs, 5,000 LRGs, and 5,000 galaxies matching the DESI BGS
sample. These were generated for rest-frame wavelengths 1,228--11,000~\AA~on a
linear 0.1~\AA~grid.  Each spectrum was normalized to have median(flux)=1 before
performing PCA so that the eigenvector variation would capture differences
between the spectra and not just normalization variations.  The mean flux was
\textit{not} subtracted prior to PCA, such that the 0th component represents the
approximate mean spectrum as an output of PCA.  Due to the large size of the
input matrix (20k by 97.7k), the PCA solution was determined iteratively using
the EMPCA algorithm\footnote{\url{https://github.com/sbailey/empca}} described
in \cite{Bailey2012-EMPCA}.

\begin{figure*}[t]
\centering
\plotone{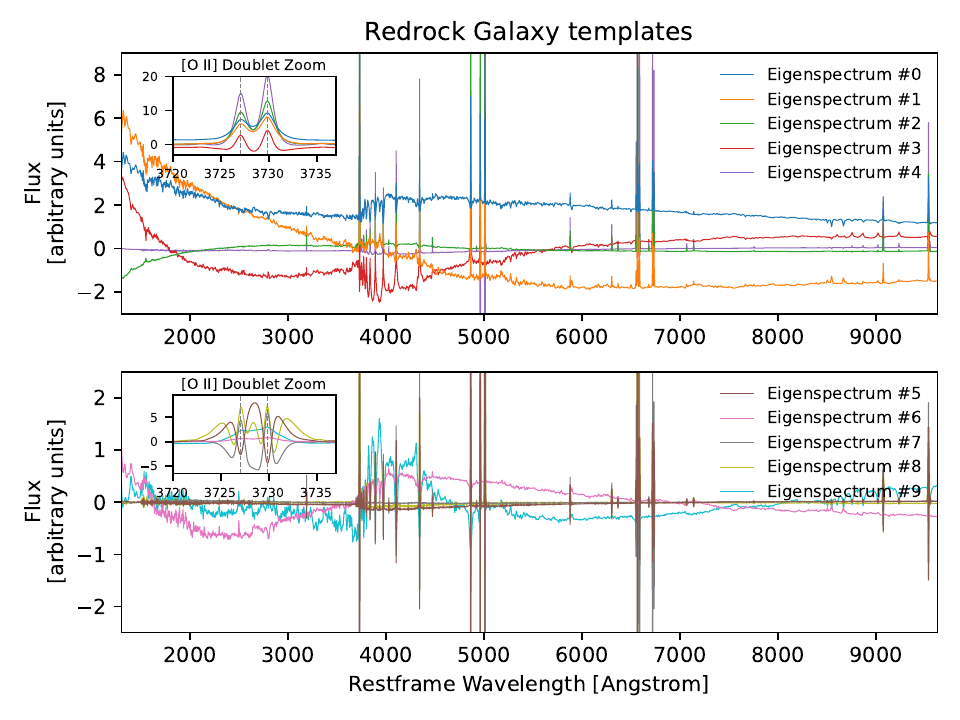}
\caption{
Galaxy eigenspectra used by Redrock, scaled for clarity. The [O~\textsc{ii}] inset illustrates how the different eigenspectra capture 
variation in emission line profile.
}
\label{fig:galaxy-templates}
\end{figure*}

The final galaxy templates use the top 10 PCA eigenvectors as shown
in Figure~\ref{fig:galaxy-templates}.
The number of eigenvector templates
was selected to balance code runtime with algorithmic performance on eBOSS
and pre-DESI simulated data, but it was not optimized in detail.
The code used to generate the galaxy PCA templates is in the \texttt{bin/rrgaltemplate} script at \url{https://github.com/desihub/redrock}.

\subsubsection{Quasar templates}
\label{sec:qso-templates}

\begin{figure*}[t]
\centering
\plotone{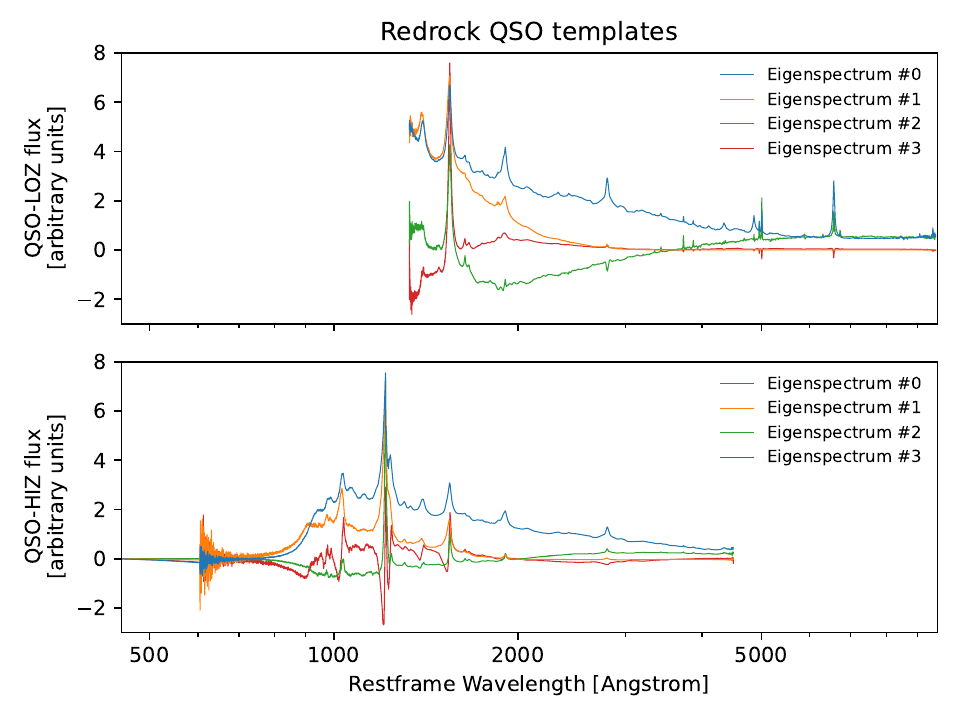}
\caption{
QSO eigenspectra used by Redrock for
Data Release 2. The eigenspectra fluxes are scaled for clarity.
}
\label{fig:qso-templates}
\end{figure*}

Quasar spectra exhibit extreme diversity in their emission and absorption properties. As such,
Redrock relies on empirically derived templates (unlike with galaxies and stars) to maximize the information content of the spectral models.

The original version of Redrock used the same QSO templates as SDSS.
\cite{Bolton2012} documents the details of their construction.\footnote{
The code used to adapt the BOSS QSO templates for Redrock is in the \texttt{bin/rrqsotemplate} script at \url{https://github.com/desihub/redrock}.}
These include 4 eigenvectors on a
$\Delta \log(\lambda)=0.0001$ logarithmic
restframe wavelength grid from 450.2~\AA~to 10399.2~\AA. 
The Lyman-series forest is present at $\lambda_{RF}<\lambda_{Ly\alpha}$ in the spectra of 
high redshift objects, leading to an overall redshift-dependent suppression of flux blueward 
of this wavelength \citep[e.g.,][]{mcquinn16}. The expected flux deficit at the redshift being
fit is accounted for by applying a wavelength-dependent correction to the templates at 
$\lambda_{RF}<\lambda_{Ly\alpha}$ following the model of \citet{calura12}.\footnote{This correction was actually incorrect for these templates, which did not incorporate an optical depth model in their derivation. See \S\ref{sec:known_issues}.}
This version of the QSO templates was used for the DESI spectroscopic pipeline
included in its Early Data Release \citep[EDR;][]{DESI2024.EDR}.

After the EDR dataset was internally frozen and released to the DESI collaboration, a new set of QSO templates was developed by \citet{Brodzeller2023.QSOTemplates}. Their \S4.1 describes the improved spectral diversity coverage of these templates relative to the EDR version, particularly for systems with dust-reddened continua or those exhibiting strong absorption.
These used a much larger set of input
SDSS DR16 quasars \citep{Lyke2020-DR16Q}, updated the Lyman-series optical depth model \citep{kamble20}, and split the sample into separate
low-redshift (LOZ, $0.05 < z < 1.6$) and high-redshift (HIZ, $1.4 < z < 7$) templates,
shown in Figure~\ref{fig:qso-templates}. 
Algorithmically, the LOZ and HIZ template sets are treated as QSO subtypes; every target is
fit with both LOZ and HIZ QSO templates over their individual redshift
ranges (see Table~\ref{tab:redshift_ranges}), and the lowest 3 QSO $\chi^2$
fits across either template set are kept for comparison with other templates.
If the true redshift lies within the overlap range of the templates
$1.4 \le z < 1.6$, typically both templates will have a minimum at the
same redshift and the one with the lower $\chi^2$ is selected.
QSOs with a true redshift just under the boundary of $z=1.4$ typically have
a good QSO-LOZ minimum fit, but the QSO-HIZ fit may extrapolate beyond its
redshift range and result in an unreliable extrapolated minimum.  This is
flagged as a bad QSO-HIZ fit and discarded and the QSO-LOZ is kept instead.
A similar effect occurs in reverse for targets just above the $z=1.6$
boundary. These cases require no QSO-specific code: an extrapolated minimum outside a template's redshift range fails the minimum-fit criteria described in \S\ref{sec:overview} and is discarded by the generic ranking of fits across templates.
This also makes it possible for future galaxy templates to be customized
for different redshift ranges.

\subsubsection{Stellar templates}
\label{sec:star-templates}

The stellar templates for Redrock were derived from the main-sequence stellar 
models from \cite{allende18}. These high-resolution ($\mathcal{R}=10,000$) models span a 
wide range of stellar effective temperature ($T_{\text{eff}}$), surface gravity 
($\log_{10} g$), and iron abundance ([Fe/H]). The basic parameters of the stellar
models used are detailed in Table~\ref{tab:stellar-model-parameters}.

\begin{table}[ht]
\centering
\begin{tabular}{l|l}
Parameter & Value Range \\
\hline
$T_{\text{eff}}$  & $3500-30,000$ K \\
$\log_{10} g$ &  $0.0 - 5.0$ cm s$^{-2}$ \\
$\left[ \text{Fe} / \text{H} \right]$ &  $-2.5-0.5$ \\
Wavelength grid &  $3000-10999.5$~\AA\ \\
\end{tabular}
\caption{
Parameters for theoretical stellar models.
}
\label{tab:stellar-model-parameters}
\end{table}

The stellar template training sample consisted of 10,000 random realizations from these models generated on a linear 0.1~\AA~wavelength
grid from 3000~\AA~to 11000~\AA.  Due to the significant spectral variation
of stars with $T_{\text{eff}}$, a single set of PCA
eigenvectors was not sufficient for accurate classification --- a small number of PCA eigenvectors were
unable to capture the variation found in stars, while a large number
of PCA eigenvectors enabled too much flexibility for unphysical fits and was additionally computationally prohibitive to use.
To accommodate the underlying stellar diversity, the training sample was split by
effective temperature $T_\mathrm{eff}$ into separate samples to generate PCA eigenvectors roughly corresponding to the physical stellar subtypes listed in Table~\ref{tab:stellar-templates}. Input spectra with $T_\mathrm{eff}$ on the boundary between subtypes
were used by both samples, thus the sum of the individual sample sizes
exceeds the original 10,000 generated.  The top 5 eigenvectors were kept
for each stellar subtype. 

A separate white dwarf PCA template set was generated using
10,000 DA and 10,000 DB subtype models \citep{koester10} on the same wavelength grid as the
other stellar spectra, also keeping the top 5 eigenspectra.
Finally, a 3-component cataclysmic variable (CV) stellar template was constructed by reformatting the \texttt{spEigenCVstar-55734.fits} model from the SDSS/BOSS
\texttt{idlspec2d} code \citep{Bolton2012}.

\begin{table}[ht]
\centering
\begin{tabular}{l|l|l|l}
Subtype & min $T_\mathrm{eff}$ & max $T_\mathrm{eff}$ & sample size \\
\hline
B  & 10,000 & 30,000 & 2,182 \\
A  &  7,000 & 10,000 & 1,466 \\
F  &  6,000 &  7,500 & 1,705 \\
G  &  5,200 &  6,000 &   955 \\
K  &  3,700 &  5,200 & 1,437 \\
M  &  3,500 &  3,700 & 3,432 \\
\end{tabular}
\caption{
Effective temperature ranges and sample size used for each Redrock stellar
template subtype.
}
\label{tab:stellar-templates}
\end{table} 

Although Redrock splits stellar templates by subtype, this is intended
primarily as an algorithmic method to accommodate broad spectral diversity
while limiting the phase space of possible template fits.
The Redrock stellar subtypes were not designed for a rigorous stellar classification.
Compared to the more detailed $T_\mathrm{eff}$ fits in \cite{Koposov_MWS_DR1_VAC}, Redrock's best fit stellar template matches only $\sim65\%$ of the time, though the neighboring template (higher- or lower-temperature range) is typically the selected one when they don't match.

\begin{figure*}[t]
\centering
\includegraphics[width=\textwidth]{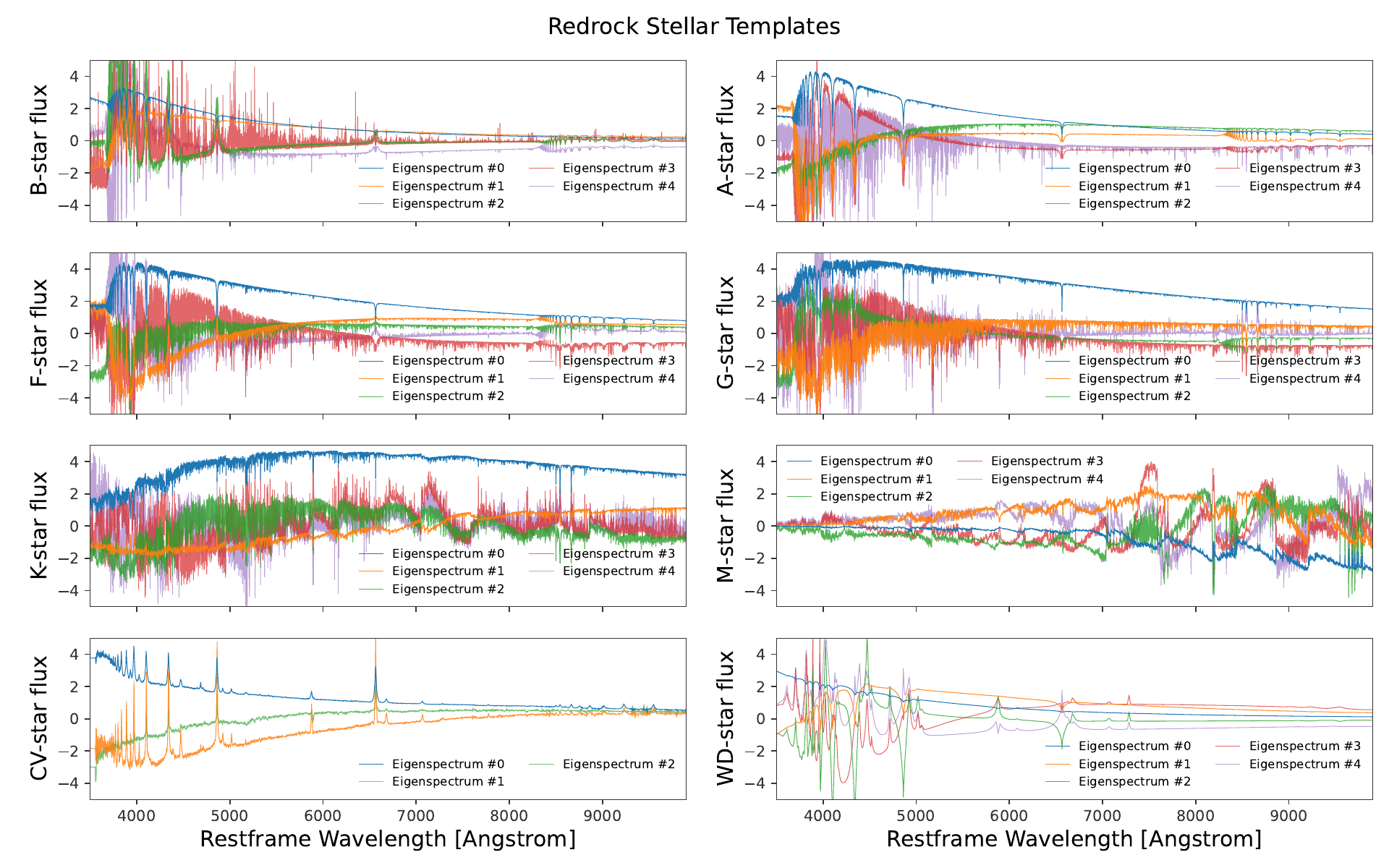}
\caption{
Redrock eigenspectra for the eight stellar subtype templates.
Flux is in arbitrary units.
}
\label{fig:star-templates}
\end{figure*}

Redrock stellar template eigenspectra are shown in
Figure~\ref{fig:star-templates}.
The code used to generate the stellar PCA templates is in the \texttt{bin/rrstartemplate} script at \url{https://github.com/desihub/redrock}.

\subsection{Archetypes}
\label{sec:archetypes}

Similar to PRIMUS \citep{cool13-primus}, eBOSS used archetype templates both with the \texttt{redmonster} software package \citep{hutchinson2016} and with Redrock. The archetypes in Redrock improved purity by avoiding unphysical solutions \citep{Ross2020.eBOSS};
these are available in a separate repository
\url{https://github.com/desihub/redrock-archetypes} tag 0.1.
The eBOSS-optimized archetypes did not work well for DESI, which observed
a broader diversity of galaxy targets, so new galaxy archetypes were developed for
DESI with details described in \cite{Anand2024.RedrockArchetypes}.
These are available as tag \texttt{v1.0.0}\footnote{\url{https://github.com/abhi0395/new-archetypes}}, though these have not yet been used in an official production run scheduled for public release.

Here, we provide a brief summary of the method. In the archetype-based approach, galaxy spectra are modeled using a set of physically motivated spectral templates, known as archetypes. These archetypes represent realistic galaxy spectral energy distributions (SEDs) and effectively capture physical features such as emission and absorption lines observed in the data. These archetypes are stored separately from the core pipeline, allowing them to be independently developed and updated.

To prevent unphysical solutions, the linear coefficients associated with these archetypes are constrained to be strictly positive using a Bounded-Value Least Squares (BVLS, \citealt{stark1995}) method. This constraint eliminates non-physical modeling artifacts, such as inverted emission or negative absorption features. Additionally, in archetype mode, we incorporate Legendre polynomials per camera to flexibly model and absorb spectrophotometric calibration issues \citep[see][for details]{guy23a,Anand2024.RedrockArchetypes}.
These include additive offsets arising from CCD amplifier bias mismatches at camera boundaries and chromatic distortions due to fiber positioning and focal plane throughput variations.
While PCA templates have the flexibility to partially absorb these effects, archetypes do not, since they are constrained to be physically meaningful, thus additional polynomial terms are helpful to absorb the unphysical effects.
The Redrock fitting algorithm allows users to define a prior on the $\chi^{2}$ contribution from the Legendre polynomial terms. This ensures that the polynomial correction primarily addresses CCD calibration issues and prevents the polynomials from mistakenly modeling intrinsic spectral features, such as the quasar power-law continuum, as calibration issues. For further details, see \citet{Anand2024.RedrockArchetypes}.

By combining physical archetypes with per-camera polynomial corrections, this approach produces redshift fits that are more physically realistic, while preserving the speed and robustness of the pipeline on DESI spectra. In the pipeline, the archetype-based method can be invoked by passing the archetypes using the \texttt{--archetypes} option to \texttt{redrock.external.desi.rrdesi}.

Another algorithmic feature that Redrock supports in archetype mode is the ability to combine multiple nearby archetypes (along with Legendre polynomial terms), rather than relying solely on the single best-fit archetype when fitting for the redshift. This approach effectively performs a ``sub-grid'' interpolation by taking the best-fitting archetype and combining it with its $N$ nearest neighbors in $\chi^2$ space. As detailed in \citet{Anand2024.RedrockArchetypes}, this method provides slightly improved $\chi^2$ values for some spectra due to the increased flexibility from a larger number of fitting coefficients. However, it also introduces significantly higher computational cost---approximately three times slower---because of the expanded template set and greater degrees of freedom. This mode can be enabled by passing the \texttt{--n-nearest NUMBER} option to \texttt{redrock.external.desi.rrdesi}. The \textit{archetype} mode was extensively tested on a large volume of DESI spectral data, and its performance is described in detail in \citet{Anand2024.RedrockArchetypes}. While Redrock performs slightly better in archetype mode, this improvement comes at the cost of longer runtime.

\section{Algorithmic Performance on DESI Data}
\label{sec:algorithm_performance}

Redrock's performance on the DESI extragalactic data set was validated as a part of DESI's Survey Validation phase \citep[SV;][]{DESI2024.SV}. \cite{lan2023} presents the results of the validation study on the galaxy (BGS, LRG, and ELG) sample, while \cite{alexander2023} reports analogous results for the QSO sample. As part of this validation process, more than 15,000 DESI spectra were visually inspected (VI) to construct high-fidelity truth catalogs.\footnote{\url{https://data.desi.lbl.gov/doc/releases/edr/vac/vi/}} These truth catalogs informed the definitions of the large-scale structure catalogs used for DESI's key cosmological analyses \citep{DESI2025.LSS,DESI.DR1.KP7,DESI.DR2.BAO}.  We compare these VI truth catalogs to the Redrock results on data from the Loa production for the upcoming DESI DR2 and repeat a similar validation analysis here. Throughout this section, true redshifts, spectral types, and quality flags are taken from the \texttt{VI\_Z}, \texttt{VI\_SPECTYPE}, and \texttt{VI\_QUALITY} columns of the VI catalogs.

The targets in the VI catalogs each received multiple exposures, up to ten times the nominal depth, which were coadded to create high S/N spectra for robust VI. The nominal depth for objects observed by DESI is 240 seconds for bright-time targets (BGS and most stellar targets) and 1000 seconds for dark-time targets (LRG, ELG, QSO). We exclusively use the Redrock results on spectra coadded to nominal depth to estimate the typical performance expected for the main survey. Classifications with non-zero \texttt{ZWARN} attributes (see \S\ref{sec:outputs}) are excluded, excepting the \texttt{SMALL\_DELTA\_CHI2} bit triggered by the first and second best solutions having $\Delta\chi^2<9$. We further exclude $\texttt{SPECTYPE=GALAXY}$ with redshift estimates $z>1.6$. Beyond this, the [O~\textsc{ii}] emission feature redshifts outside of DESI wavelength coverage, and therefore reliable galaxy classifications are not expected \citep[e.g.,][]{LRG.TS.Zhou.2023,ELG.TS.Raichoor.2023}. 
Finally, a high-quality VI classification is required, defined as \texttt{VI\_QUALITY > 2.5}. 

We first measure redshift recovery rate, sample completeness, and sample purity for each of DESI's extragalactic target classes. Redshift recovery rate $R$ is defined as
\begin{equation}
\label{eq:redshift_recovery}
    R = \frac{N(dz<dz_\mathrm{limit}\ \&\ \texttt{IS\_OBJ})}{N(\texttt{IS\_OBJ})}
\end{equation}
where \texttt{IS\_OBJ} is either true galaxy type or true quasar type and
\begin{equation}
\label{eq:dz}
dz = \frac{|z_{VI} - z_{redrock}|}{1 + z_{VI}}.
\end{equation}
For galaxies, it measures the percentage of true galaxies that receive Redrock redshift estimates with $dz<0.0033$ (equivalent to $dv\approx1000$~km~s$^{-1}$).
For quasars, the redshift recovery rate uses $dz_\mathrm{limit}=0.01$ ($dv \approx 3000$~km~s$^{-1}$). The relaxed criterion relative to galaxies accommodates redshift uncertainty introduced by broad emission lines \citep[e.g.,][]{hewett2010}.

Sample completeness $C$, on the other hand, additionally requires a correct \texttt{SPECTYPE} assignment from Redrock.
\begin{equation}
\label{eq:completeness}
    C = \frac{N(dz<dz_\mathrm{limit}\ \&\ \texttt{IS\_OBJ}\ \&\ \texttt{SPECTYPE}=\texttt{OBJ})}{N(\texttt{IS\_OBJ})}
\end{equation}
The sample completeness is always lower than the redshift recovery rate, due in part to the ambiguity introduced by AGN activity. This is discussed more in \S\ref{sec:known_issues}.

Finally, sample purity $P$ measures the percent of \texttt{SPECTYPE=GALAXY} classifications that are true galaxies with $dz<0.0033$, and similarly for \texttt{SPECTYPE=QSO} with $dz<0.01$.
\begin{equation}
\label{eq:purity}
   P = \frac{N(dz<dz_\mathrm{limit}\ \&\ \texttt{IS\_OBJ}\ \&\ \texttt{SPECTYPE}=\texttt{OBJ})}{N(\texttt{SPECTYPE}=\texttt{OBJ})}
\end{equation}
The results are summarized in Table~\ref{tab:rr_performance}.

\begin{figure}
\centering
\includegraphics[width=\columnwidth]{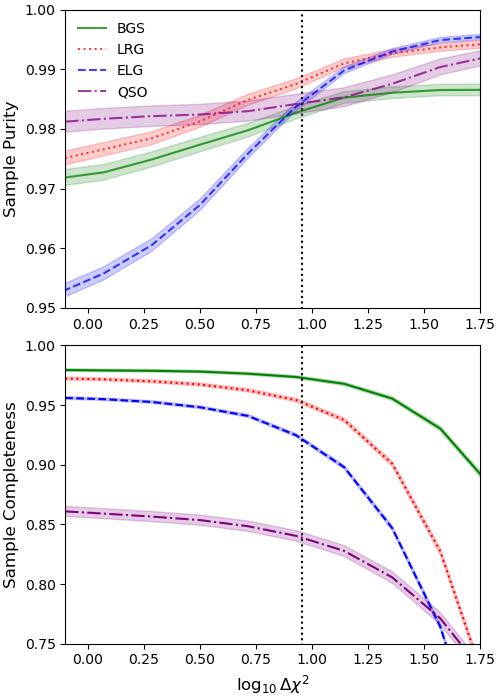}
\caption{The dependence of sample purity and completeness on $\Delta\chi^2$, the difference in $\chi^2$ between the first and second best Redrock solution, for each target class. The vertical dashed line at $\Delta \chi^2 = 9$ indicates the threshold to trigger the \texttt{SMALL\_DELTA\_CHI2} \texttt{ZWARN} bit. }
\label{fig:pc_vs_deltachi2}
\end{figure}

Figure~\ref{fig:pc_vs_deltachi2} illustrates the dependence of purity and completeness on the \texttt{DELTACHI2} parameter. As illustrated in the figure, $\Delta\chi^2<9$ is an adequate threshold to generically indicate lower confidence classifications. This parameter, however, can be tuned for specific use cases as is done with the DESI large-scale structure catalogs \citep[LSS;][]{DESI2025.LSS}. The LSS catalogs use optimized selection cuts for each target class to meet the scientific requirements for DESI \citep{abareshi22a}. For example, the BGS sample uses a $\Delta\chi^2 >40$ and requires $z<0.6$ \citep{BGS.TS.Hahn.2023} while the LRG sample is defined by $\Delta\chi^2 >15$ and $0.4<z<1.1$ \citep{LRG.TS.Zhou.2023}. The ELG and QSO target classes were more difficult to optimize owing to the nature of the sources, requiring additional classifiers to construct the LSS catalogs \citep[see][for complete details on these classifiers]{ELG.TS.Raichoor.2023,QSO.TS.Chaussidon.2023}. The DESI LSS redshift purity rates are included in Table~\ref{tab:rr_performance} for comparison to the Redrock baseline.
\begin{table*}
\centering
\begin{tabular}{l|l|l|l|l}
Target & Redshift      & Sample       & Sample  & LSS Redshift\\
Class  & Recovery Rate & Completeness & Purity & Purity \\
\hline
BGS    &  98.2\%  &  98.0\%   &  96.9\%   & $>99.9$\% \\
LRG    &  97.9\%  &  97.4\%   &  97.3\%   & 99.6\% \\
ELG    &  95.9\%  &  95.9\%   &  94.7\%  & $>99.6$\% \\
QSO    &  92.6\%  &  86.3\%   &  98\%  & 99\% \\
\end{tabular}
\caption{The redshift recovery rate, completeness, and purity achieved by Redrock for the primary extragalactic target classes in DESI. These metrics are measured for classifications with $\texttt{ZWARN} = [0,4]$. Restricting to $\texttt{ZWARN} = 0$ results in slightly lower ($\sim$1-3\%) redshift recovery rates and completeness, but can improve purity (notably for ELGs). The values in the LSS Redshift Purity column are those reported by \cite{lan2023} and \cite{alexander2023}. The LSS samples use additional quality cuts to maximize success in the redshift ranges used for cosmological analyses.}
\label{tab:rr_performance}
\end{table*} 

\begin{figure}
\centering
\includegraphics[width=\columnwidth]{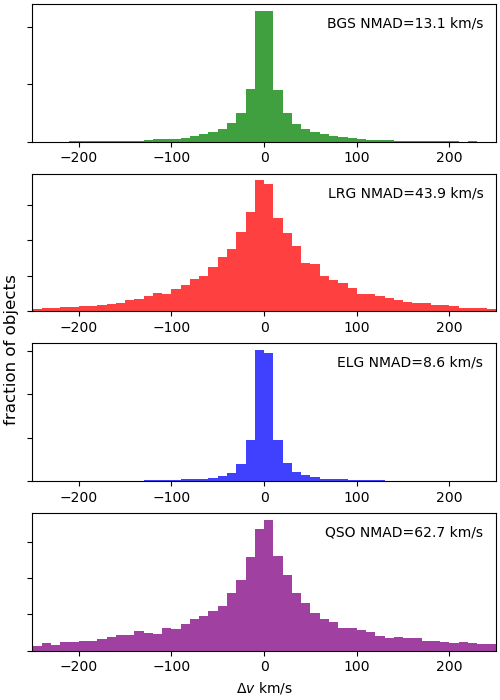}
\caption{Random redshift error measured from the distribution of redshift-pair offsets for each DESI extragalactic target class.
}
\label{fig:redshift_precision}
\end{figure}

The random error on Redrock redshift estimates is determined by comparing the results for the same object in multiple exposures. The velocity offset between redshift pairs $(i,j)$ is calculated as 
\begin{equation}
    dv_{ij} = c (z_i - z_j) / (1+z_i)
\end{equation}
We then use the normalized median absolute deviation (NMAD) of this distribution as an estimator for random redshift error, defined as:
\begin{equation}
    \text{NMAD} = \frac{\text{MAD}\times1.4828}{\sqrt{2}}
\end{equation}
This statistic is chosen over the standard deviation because it is more robust against catastrophic failures. The velocity offset distribution for each DESI target class is shown in Figure~\ref{fig:redshift_precision}, with the corresponding NMAD values annotated.
ELG targets are expected to have the lowest random redshift error (8.6~km~s$^{-1}$) due to their sharp emission lines,
while QSOs have the largest random errors (62.7~km~s$^{-1}$) due to their broad lines.
All target classes exhibit non-Gaussian tails in their repeat-observation $dv$ distributions.

\begin{figure}
\centering
\includegraphics[width=\columnwidth]{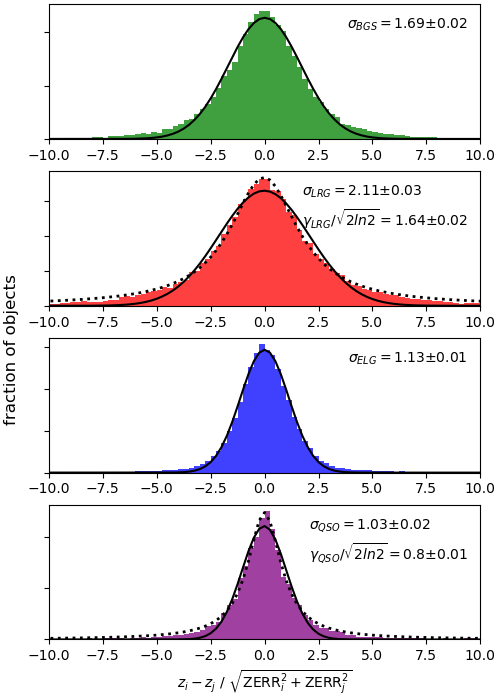}
\caption{Accuracy of the estimated redshift error \texttt{ZERR} from Redrock for DESI's four primary extragalactic target classes. The solid lines represent the best-fitting Gaussian functions. The dotted lines represent the best-fitting Lorentzian functions, if applicable to the given distribution.}
\label{fig:zerr_accuracy}
\end{figure}

We next use the redshift pairs to estimate the accuracy of Redrock's \texttt{ZERR} parameter. If \texttt{ZERR} is an accurate representation of redshift uncertainty, then the distribution of $(z_i - z_j)/\sqrt{\text{ZERR}_i^2 + \text{ZERR}_j^2}$ will resemble a Gaussian with $(\mu,\sigma)=(0,1)$. Figure~\ref{fig:zerr_accuracy} shows these distributions for the four DESI extragalactic target classes. A Gaussian is fit to each of these distributions, and the standard deviation $\sigma$ is annotated. The LRG and quasar samples exhibit significant tails, so we additionally fit them with a Lorentzian distribution and report the best-fit half-width at half-maximum, scaled to represent the standard deviation. The Lorentzian tails for the LRGs and QSOs are driven by pairs that both have median SNR less than 1.5, indicating the error model is less reliable in this regime.

We conclude that \texttt{ZERR} accurately reflects redshift uncertainty for the quasar population; however, it is underestimated for BGS by $\sim$70\%, and for ELGs by $\sim$10\%. Restricting to pairs with SNR~$>1.5$, we find \texttt{ZERR} for LRGs is underestimated as well by about 65\%.
For comparison, BOSS underestimated galaxy errors by 20--30\% and QSOs by a factor of 2 \citep{Bolton2012}.

Finally, we use the Milky Way Survey (MWS) catalog produced by the \texttt{RVSpecFit}\footnote{\url{https://github.com/segasai/rvspecfit}} (RVS) software \citep[][]{koposov11,li19} to check performance on stellar spectra. In short, RVS is run on the subset of spectra in the Loa production that are suspected stellar candidates (i.e., objects with MWS targeting bits or \texttt{SPECTYPE=STAR} from Redrock). The output catalog contains maximum likelihood estimates of various stellar parameters, including radial velocity accurate to approximately 1~km~s$^{-1}$. RVS also provides a boolean \texttt{SUCCESS} flag per object to indicate whether the results are reliable. Complete details on the RVS software with respect to the DESI MWS are provided by \cite{koposov24, cooper23}.

\cite{koposov24} demonstrated that requiring RVS \texttt{SUCCESS=True} and \texttt{SPECTYPE=STAR} produces a relatively clean stellar sample. We use the RVS radial velocity measurements from this sample to estimate the radial velocity accuracy of the stellar templates.
For each object, the offset $dv_{rad}$ is computed as
\begin{equation}
dv_{rad} = v_{RVS} - c*z_{redrock} .
\label{eq:rad_velocity}
\end{equation}
The distribution indicates fair agreement between the two estimators, with a mean of $0.14$~km~s$^{-1}$ and a standard deviation of $5.3$~km~s$^{-1}$. We do not validate other stellar parameters since the stellar templates are not intended for robust subtype classifications (\S\ref{sec:star-templates}). In addition, Redrock imposes a limit of $|cz| < 600$~km~s$^{-1}$ for \texttt{SPECTYPE=STAR}, limiting its accuracy for high-velocity stellar targets. Owing to the detailed physical modeling employed for the MWS catalog, that catalog is recommended for scientific analyses of the DESI stellar sample. 

\section{Known Issues}
\label{sec:known_issues}

\subsection{Physical Constraints}

PCA templates provide a way to flexibly describe a spectrum with a small
number of eigenvectors that can be fit with fast linear algebra, but they
can also be overly flexible and result in unphysical fits.  For example,
a small flux calibration offset between two spectrograph arms could be fit
as an inverse 4000\AA~break, or a mis-subtracted sky line could be fit with
a negative emission line.  Optional archetypes and a $\chi^2$ penalty on negative
\oii~flux are current methods by which Redrock mitigates unphysical fits,
but more could be done.  Ongoing work includes adding support for Non-Negative Matrix Factorization (NMF) templates \citep{Zhu2016-NMF}, improving the
negative-emission line penalties, and exploring templates as an interpolation
of models on a grid of physical parameters similar to the standard star fits
in \cite{guy23a}~\S4.8.

\subsection{Template Degrees of Freedom}
As described in \S\ref{sec:overview}, Redrock selects the best
classification using total $\chi^2$ without accounting for the
different number of coefficients in each template set. This mildly
favors the 10-coefficient galaxy templates for the default DESI template set.
To quantify the impact, we re-ranked the Redrock solutions by reduced $\chi^2$ instead of total $\chi^2$ for a random 30\% of DESI DR1 tiles. An estimated 0.3\% of target spectra would receive a different \texttt{SPECTYPE}; however, these changes are dominated by spectra already flagged with
\texttt{SMALL\_DELTA\_CHI2}. Among spectra with \texttt{ZWARN=0}, less than 0.1\% would receive a different \texttt{SPECTYPE}. The overwhelming majority of these changes would be from \texttt{GALAXY} to \texttt{QSO-LOZ} or to a stellar subtype at effectively the same redshift.

\subsection{Galaxies with AGN}

The default Redrock templates do not include Active Galactic Nucleus (AGN)
galaxies in their training sample.
These galaxies typically have a mix of broad and narrow emission lines and
thus could be fit at the same redshift by either galaxy or QSO templates
and assigned to either \texttt{SPECTYPE=QSO} or \texttt{SPECTYPE=GALAXY}.  Future work could expand the galaxy templates to include
AGN-like broad line features,
perhaps as a separate galaxy template subtype. The current lack of an AGN subtype does not pose any issues for LSS studies in DESI however since it would simply be an additional subclassification, rather than meaningfully impacting redshifts. AGN-like galaxies are predominantly assigned a correct redshift even if assigned \texttt{SPECTYPE=QSO}. Further, achieving high completeness for the QSO spectral type classification has proven more difficult than with galaxy spectral type \citep{alexander2023} so the potential effect of some current \texttt{SPECTYPE=QSO SUBTYPE=LOZ} QSOs
being instead classified as \texttt{SPECTYPE=GALAXY SUBTYPE=AGN} would need to be carefully assessed.

\subsection{Lyman Series Optical Depth}

The original default QSO templates used for the DESI Early Data Release \citep[EDR;][]{DESI2024.EDR} and the QSO-HIZ v1.0 templates used for DESI Data Release 1 \citep[DR1;][]{DESI2025.DR1} both had a mismatch with the Redrock code for how the mean
Ly$\alpha$ absorption should be handled (in the templates vs.~in the code). 
This led to an increasing redshift bias for QSOs at $z>2$ \citep[e.g.,][]{wu23}. The average redshift bias ranged from $-109 \pm 17$~km~s$^{-1}$ at $z\sim2$ to $-358 \pm 41$~km~s$^{-1}$ at $z>2.5$ in EDR \citep{Bault24a.QSOz} and $9 \pm 29$~km~s$^{-1}$ to $-84 \pm 31$~km~s$^{-1}$ in DR1 \citep{Brodzeller2023.QSOTemplates}.
This was corrected in the QSO-HIZ v1.1 templates used for the Loa production to be
released with DESI Data Release 2.  Corrected QSO redshifts using
QSO-HIZ v1.1 templates are included in DESI DR1 as a ``zlya'' value-added catalog\footnote{\url{https://data.desi.lbl.gov/doc/releases/dr1/vac/zlya/}} \citep[constructed for the BAO analysis presented by][]{DESI2025.Lya}.

\subsection{Quasar Diversity}

The current QSO templates have 4 basis vectors for each of the HIZ and LOZ
subtypes. This is sufficient for the primary DESI QSO sample, but does not capture the full diversity of QSOs with significant Broad Absorption Lines (BALs), highly reddened QSOs, or QSOs whose properties lie well outside of
the standard SDSS/BOSS and DESI QSO target selection \citep{Yang2023.QSO, Juneau2024.QSO}.  This could be improved by training a new set of QSO templates directly on DESI data, which would also have the added benefit of improving QSO template resolution. However, initial studies have found it surprisingly difficult to improve upon the current templates without inadvertently adding too much flexibility causing a performance degradation through false-positive fits of galaxy data.
For DESI, the current Redrock QSO performance is augmented through additional
QSO-specific algorithms using a neural net \citep{Busca18.QuasarNet, Farr2020.QuasarNet, Green2025.QuasarNet}
and fitting for the \mgii\ emission line \citep{QSO.TS.Chaussidon.2023}. These are run
separately from Redrock and are used in combination with Redrock when generating
the DESI QSO catalogs.

\subsection{Exceptionally Narrow Emission Lines}

The redshift resolution of the current galaxy templates (Table~\ref{tab:redshift_ranges}) corresponds to a spacing between the redshift grid points of $dz=0.0007$ at $z=0$ and  $dz=0.0019$ at $z=1.7$. This resolution was chosen to balance performance with compute time and output file size. However, in rare instances where a spectrum exhibits only one atypically narrow emission line, this resolution is low enough that it can step over this feature. We tested a galaxy redshift grid with half the current step size, finding only a handful of cases where the solution improved while many of the changed solutions now confused sky lines for emission lines. We therefore concluded the default step size is preferred for the current DESI instrumentation and target classes. A finer grid, or one with constant steps in $z$ instead of steps uniform in $\log(1+z)$, should be considered if Redrock is used for higher-redshift galaxy classification.

\subsection{Increased Catastrophic Failure Rate for $z>1.5$ DESI ELGs}

As galaxy redshifts increase, fewer emission lines are available for fitting until eventually only the [O~\textsc{ii}] emission line remains visible to the DESI spectrographs. At $z>1.5$, the [O~\textsc{ii}] line coincides with regions of strong sky emission and
sky-subtraction residuals can persist even in high S/N spectra. These residuals impact Redrock's ability to obtain secure redshifts for these galaxies, causing the rate of catastrophic redshifts for ELG targets to noticeably rise. \S7 of \citet{ELG.TS.Raichoor.2023} explores the impact of sky residuals on ELG classification in detail.

\section{Computational Performance}
\label{sec:computational_performance}

\begin{table}
\begin{tabular}{l|l|l}
Step & CPU-only & CPU+GPU \\
     & [seconds] & [seconds] \\
\hline
I/O                       &   24.8 &   31.5 \\
Rebin templates           &    2.6 &    1.0 \\
GPU overhead              &    0.0 &    1.6 \\
Initial redshift scan     &  274.5 &   34.2 \\
Rescan + fit minima       &   66.7 &  178.0 \\
Finalize results          &   10.6 &   12.2 \\
\hline
Total                     &  379.2 &  258.5 \\
\end{tabular}
\caption{Timing results for processing 5000 DESI spectra (one exposure)
on a NERSC Perlmutter GPU node. }
\label{tab:timing}
\end{table}

Table~\ref{tab:timing} shows the runtime for processing 5000 DESI spectra
(one exposure) using a GPU node on the Perlmutter system at the
National Energy Research Scientific Computing Center
(NERSC)\footnote{\url{https://www.nersc.gov/systems/perlmutter/}}; these nodes
have a 64-core AMD EPYC CPU and four NVIDIA A100 GPUs. Other GPUs, \eg, those in Apple-silicon laptops, are not currently supported,
and Redrock runs in CPU-only mode on such systems. Timing numbers are
shown for using just the CPU and for using the CPU and the GPUs
together.\footnote{Perlmutter also has CPU-only nodes with 128 cores;
the numbers shown here are for using just the 64-core CPU on the GPU nodes.}
The ``Initial redshift scan'' step is dramatically faster on GPUs than on
CPUs. This comes at a cost of internal data structure complexity and bookkeeping
moving data between the GPUs and CPUs which causes ``Rescan + fit minima'' step to be \textit{slower} on the GPUs, though the combined results
are still faster when using GPUs.

One exposure of processed DESI data has 2.1 GB of input coadded spectra and 1.3 GB
of Redrock outputs.\footnote{The primary Redrock output table with 
classifications and redshifts is small (5.4 MB); the output size is dominated
by the optional details file containing the full $\chi^2$~\vs~$z$ scan for
every template for every target and additional details for each minimum fit.}
It takes less than 5 minutes to process all 5000 spectra for a DESI exposure
on these GPU nodes.  For the final DESI survey of $\sim$60 million targets,
this corresponds to $\sim$1000 node-hours on Perlmutter. For reference, this accounts for $\sim$13\% of the node-hours required by the full pipeline from calibrations through to redshifting.

\section{Output files}
\label{sec:outputs}

Redrock is a Python software package with experiment-agnostic algorithms, along with separate wrapper scripts for experiment-specific input and output formats.  This section documents
the formats output by the DESI wrapper (\texttt{redrock.external.desi.rrdesi}),
while noting that other experiments could use the same underlying algorithms
with different input and output formats (as eBOSS did in the
\texttt{redrock.external.rrboss} wrapper).

The primary output file is a multi-HDU (header data unit) FITS format
file \citep{Pence2010-FITS}, \texttt{redrock-*.fits}, summarized in Table~\ref{tab:redrock_fits}.
The REDSHIFTS HDU contains the Redrock classification and redshift output,
while the other HDUs propagate input metadata about the targets and spectra
so that the files are usefully self-contained.  Further details about the
REDSHIFTS HDU are listed in Table~\ref{tab:redrock_redshifts_hdu}.
Columns ZERR, ZWARN, CHI2, COEFF, NPIXELS, NCOEFF, and DELTACHI2 use
more bits than necessary (int64 or float64) and may be stored as
int32 or float32 in future DESI data releases.

\begin{table*}
\begin{tabular}{l|l|l}
HDU & & \\
Number & Name & Description \\
\hline
0 & (not set) & Empty \\
1 & REDSHIFTS & Best fit classification and redshift for each input spectrum \\
2 & FIBERMAP & Input target metadata, \eg~fluxes and targeting bits \\
3 & EXP\_FIBERMAP & Exposure-specific target metadata, \eg~fiber positions on-sky \\
4 & TSNR2 & Template signal-to-noise ratios squared
\end{tabular}
\caption{Format summary of the DESI-specific \texttt{redrock*.fits} output file.
The \texttt{REDSHIFTS} HDU is the only new information added by Redrock; the
other HDUs are propagated from the input coadded spectra files.
For further details see
\url{https://desidatamodel.readthedocs.io/en/latest/DESI_SPECTRO_REDUX/SPECPROD/tiles/GROUPTYPE/TILEID/GROUPID/redrock-SPECTROGRAPH-TILEID-GROUPID.html}
}
\label{tab:redrock_fits}
\end{table*}

\begin{table*}
\begin{tabular}{l|l|l}
Column & Type & Description \\
\hline
TARGETID   & int64       & Unique target identifier \\
Z          & float64     & Redshift \\
ZERR       & float64     & Estimated redshift error \\
ZWARN      & int64       & Bitmask of fit warnings; 0=good \\
CHI2       & float64     & $\chi^2$ of best fit template to data \\
COEFF      & float64[10] & PCA coefficients of best fit \\
NPIXELS    & int64       & Number of input pixels (wavelengths) included in fit \\
SPECTYPE   & byte6       & Spectral classification \\
SUBTYPE    & byte20      & Spectral subtype \\
NCOEFF     & int64       & Number of PCA coefficients used \\
DELTACHI2  & float64     & difference in $\chi^2$ to next best fit
\end{tabular}
\caption{Format of the REDSHIFTS binary table HDU for the DESI-specific
\texttt{redrock*.fits} output file.}
\label{tab:redrock_redshifts_hdu}
\end{table*}

Redrock encodes warnings about the quality of the fit in the ZWARN bitmask
column of the REDSHIFTS HDU of the Redrock output FITS file.
Table~\ref{tab:zwarn} lists the bits defined in Redrock.
Bits 0-9 inherit their name and meaning from SDSS (see \cite{Bolton2012}),
though bits 4 (MANY\_OUTLIERS), 6 (NEGATIVE\_EMISSION), and 8 (BAD\_TARGET)
are currently not implemented in Redrock.

\begin{table*}
\begin{tabular}{l|l|l}
Bit number & Name & Description \\
\hline
0  & SKY                & sky fiber \\
1  & LITTLE\_COVERAGE   & too little wavelength coverage \\
2  & SMALL\_DELTA\_CHI2 & chi-squared of best fit is too close to that of second best \\
3  & NEGATIVE\_MODEL    & archetype fit model is negative \\
4  & MANY\_OUTLIERS     & UNUSED; More than 5\% of data points are $>5\sigma$ from best model fit \\
5  & Z\_FITLIMIT        & chi-squared minimum at edge of the redshift fitting range \\
6  & NEGATIVE\_EMISSION & UNUSED; a QSO line exhibits negative emission \\
7  & UNPLUGGED          & the fiber was unplugged/broken, so no spectrum obtained \\
8  & BAD\_TARGET        & UNUSED; catastrophically bad targeting data \\
9  & NODATA             & No data for this fiber, \ie, all input wavelengths are masked or have ivar=0 \\
10 & BAD\_MINFIT        & Bad parabolic fit to the $\chi^2$ minimum \\
11 & POORDATA           & Poor input data quality but try fitting anyway \\
12-15 & Reserved & Reserved for future use by Redrock \\
16-23 & Reserved & Reserved for future use by experiment-specific post-Redrock processing \\
24-31 & Reserved & Reserved for future use by Redrock
\end{tabular}
\caption{ZWARN bits defined by Redrock.}
\label{tab:zwarn}
\end{table*}

Bits 16--23 are reserved for experiment-specific post-Redrock processing, that is, Redrock never sets these bits, so that the ZWARN bitmask can be propagated to downstream files while setting those bits for other purposes.  DESI uses these for augmenting the Redrock output with
quality assurance (QA) warning flags in the ``merged target list,'' downstream
from the Redrock output itself \citep{Schlafly2023.DESISurvey}.
Bits 12--15 and 24--31 are currently not used, but may be used for additional
Redrock warning flags in the future.

More detailed outputs are available in HDF5
format\footnote{\url{https://www.hdfgroup.org/HDF5/}} in the \texttt{rrdetails-*.h5} files,
including the best 3 redshift fits per template SPECTYPE for each TARGETID
(\texttt{zfit/TARGETID/zfit} table); 
and the $\chi^2$~\vs~$z$ per target per template SPECTYPE
(\texttt{zscan/SPECTYPE:::SUBTYPE/zchi2} and \texttt{redshifts} arrays).
HDF5 supports a nested hierarchy of datasets as shown in Figure~\ref{fig:rrdetails}.
For example, \texttt{zscan/QSO:::HIZ/redshifts} contains the array of redshifts scanned
for the QSO HIZ template, and \texttt{zscan/QSO:::HIZ/zchi2[i]} gives the $\chi^2$ fit
values for target \texttt{targetids[i]} at each of those redshifts.

\begin{figure*}
\noindent\hrulefill
\begin{verbatim}
rrdetails HDF5 file structure:
    targetids: int64 array[numtargets] of input unique target identifiers
    zscan:
        SPECTYPE or SPECTYPE:::SUBTYPE (e.g. "GALAXY" or "QSO:::HIZ"):
            redshifts: 1D array[numredshifts] of redshifts fit for this SPECTYPE
            zchi2: 2D array[numtargets,numredshifts]
                   chi2 values for this SPECTYPE and this redshift
            zcoeff: 3D array[numtargets,numredshifts,numcoeff]
                    best-fit PCA coefficients for this SPECTYPE at this redshift
            penalty: 2D array[numtargets,numredshifts]
                     chi2 penalty values
    zfit:
        TARGETID (str, not int):
            zfit: table of top 3 redshift fits per SPECTYPE for this TARGETID
\end{verbatim}
\noindent\hrulefill
\caption{Summary of the nested structure of Redrock details HDF5 files.}
\label{fig:rrdetails}
\end{figure*}

The final DESI Redrock output file \texttt{rrmodel-*.fits}
contains the pre-rendered best-fit templates for each target.
This file mimics the format of the input spectra files such that the
\texttt{i}th row of the model data corresponds to the \texttt{i}th input spectrum.

\section{Code}
\label{sec:code}

Redrock code is available from \url{https://github.com/desihub/redrock}
using an open source BSD 3-clause license\footnote{\url{https://opensource.org/license/BSD-3-clause/}}.
Templates and Archetypes are kept in separate repositories so that they
can be developed, versioned, and tagged independently of the code using them.
The templates used by DESI are available at \url{https://github.com/desihub/redrock-templates}, and archetypes at \url{https://github.com/desihub/redrock-archetypes}.
The DESI data processing runs for the Early Data Release used
Redrock tag 0.15.4 and templates tag 0.7.2.  Data Release 1 used
Redrock tag 0.17.0 and templates tag 0.8.
Neither used archetypes, which were still under development
at the time of those data processing runs.

Redrock is written in Python, leveraging
NumPy\footnote{\url{https://numpy.org}} \citep{harris20a.numpy} and
SciPy\footnote{\url{https://scipy.org}} \citep{virtanen20a.scipy}
for efficient compiled-code wrappers
to matrix and linear algebra operations.  It additionally uses
Numba\footnote{\url{https://numba.pydata.org}} \citep{lam2015numba}
for just-in-time (JIT) compilation of the custom trapezoidal integration used for template rebinning;
prior to using Numba this step dominated the runtime.
NVIDIA Graphics Processing Units (GPUs) are supported through the use of
CuPy
which provides a NumPy compatible
API to GPU algorithms as well as JIT compilation of custom CUDA kernels.
Redrock supports parallelism through Message Passing Interface (MPI)
mpi4py \citep{mpi4py} or with Python multiprocessing.
Overall, Redrock is designed to be capable of
running on a laptop with no GPU or MPI installed, while leveraging MPI
and GPUs on a supercomputer for large-scale data processing.

The primary Redrock code is experiment-agnostic, with anything DESI-specific
isolated to \texttt{redrock.external.desi}, \eg~file I/O format wrappers and
interpreting what is meant by the ``POORDATA'' ZWARN bit
(see \S\ref{sec:outputs}).  Keeping the templates separate from the code
also facilitates reusing the Redrock code with new templates optimized for
different types of galaxies, \eg, DESI has explored custom templates for
Lyman-$\alpha$ emitter (LAE) galaxies in preparation for future observing
programs.

The code structure of experiment-specific wrappers on top of an 
experiment-agnostic toolkit, plus making the code open-source from the start,
enabled Redrock to be used by eBOSS well before DESI began operations.  This
benefited both collaborations --- eBOSS gained a redshift fitter that
outperformed their default code, while at the same time they contributed
improvements and real data testing that accelerated DESI's readiness for
operations.

\subsection{Template file format}

Redrock uses the environment variable \texttt{\$RR\_TEMPLATE\_DIR}
to define a directory containing templates to use.  This is kept separate
from the Redrock code itself to enable future projects to develop templates
customized for their needs while still using the same Redrock code.
Templates used by DESI include galaxy,
QSO, and star templates.  By default, Redrock will use the template
files listed in \texttt{templates-default.txt}, though it can optionally
use other files to define which templates to load, \eg,
\texttt{templates-fuji.txt} defines the templates used in the
``Fuji'' production included in the DESI Early Data Release \citep{DESI2024.EDR},
\texttt{templates-iron.txt} defines the set used for DESI Data Release 1 \citep{DESI2025.DR1},
and \texttt{templates-loa.txt} defines the set used for future DESI Data Release 2.\footnote{
DESI data releases can include multiple production runs that potentially use different templates,
thus the template files are named after the production run (fuji, iron, loa), not the data release (edr, dr1, dr2).}

The templates are stored in a FITS-format \citep{Pence2010-FITS} file Header Data Unit (HDU) 0 with \texttt{EXTNAME=BASIS\_VECTORS}.  The templates are a 2D array of
number of templates by number of wavelengths.
Header keywords \texttt{CRVAL1} and \texttt{CDELT1} define the starting
wavelength and wavelength step, in Angstroms.
If present, keyword \texttt{LOGLAM=1} indicates that the wavelength grid
is expressed as $\log_{10}(\lambda)$ instead of a linear wavelength grid.
Keywords \texttt{RRTYPE} and \texttt{RRSUBTYP} define the spectral type
(classification) and subtype (blank if not applicable).
Authors of new templates are strongly encouraged to include a
\texttt{VERSION} keyword for traceability.  Note that \texttt{VERSION}
refers to the version of the individual template, not the tagged version
of the Redrock code nor any tags of the redrock-templates repository.

The templates provided with Redrock follow the naming convention
\texttt{rrtemplate-RRTYPE-RRSUBTYP-VERSION.fits}.  This allows downstream code to reconstruct the matching template name using the \texttt{RRTYPE}, \texttt{RRSUBTYPE}, and \texttt{VERSION} metadata in the Redrock output files, though this naming scheme is not required by the Redrock code.

Other HDUs are allowable but ignored.  For example, the
\texttt{rrtemplate-GALAXY-None-v2.6.fits} templates include a second
HDU recording the coefficients of the PCA fits to the training data used
to generate the 10 PCA templates in HDU 0 \texttt{BASIS\_VECTORS}.

\section{Conclusions}
\label{sec:conclusions}

Redrock is a PCA-based spectral classifier designed for high-accuracy, automated spectral typing 
and redshift estimation. It is tailored to the needs of the DESI survey, but remains generic enough 
to be used by other spectroscopic surveys (e.g., SDSS). One of the key features of Redrock is that
it efficiently leverages the full spectrum for classification, rather than relying on a specific 
subset of emission lines. The Redrock software package includes PCA template sets for several
stellar subtypes, galaxies, and quasars. While fitting, Redrock convolves its templates with the per-spectrum resolution matrix, accurately modeling the non-Gaussian, wavelength-dependent instrumental resolution while preserving the uncorrelated error model of the data.
The performance of Redrock with this template suite on
a high-fidelity truth-labeled catalog of DESI spectra demonstrates its efficacy, achieving 
excellent redshift recovery rates for all extragalactic target classes and accurate stellar 
radial velocities. Redrock has proven to be an accurate, robust, and 
computationally efficient core component of the DESI spectroscopic pipeline, capable of
classifying and fitting redshifts for the survey's diverse target populations. Its open design allows
for continuous improvement and application to future spectroscopic endeavors.

\section*{Data Availability}
The plots in this paper are based upon data released as part of DESI Data Release 1 (DR1), except the performance analysis of \S\ref{sec:algorithm_performance} which uses the forthcoming DR2.  The code to reproduce
the figures is available at
\url{https://github.com/desihub/redrockpaper}. 

\begin{acknowledgments}

SB is the developer and current maintainer of the Redrock software. AB developed the
default DESI QSO PCA templates and led the analysis presented in \S\ref{sec:algorithm_performance}. JM developed the default DESI galaxy PCA templates and conducted various efficiency studies.
AA developed the DESI-based archetypes and validated the performance of Redrock's archetype mode on DESI data. CW ported Redrock to GPUs.
TK implemented the MPI parallelization. The authors thank Hélion du Mas des Bourboux for early Redrock development as part of eBOSS.

JM gratefully acknowledges funding support for this work from the
U.S. Department of Energy, Office of Science, Office of High Energy
Physics under Award Number DE-SC0020086.

This material is based upon work supported by the U.S. Department of Energy (DOE), Office of Science, Office of High-Energy Physics, under Contract No. DE–AC02–05CH11231, and by the National Energy Research Scientific Computing Center, a DOE Office of Science User Facility under the same contract. Additional support for DESI was provided by the U.S. National Science Foundation (NSF), Division of Astronomical Sciences under Contract No. AST-0950945 to the NSF’s National Optical-Infrared Astronomy Research Laboratory; the Science and Technology Facilities Council of the United Kingdom; the Gordon and Betty Moore Foundation; the Heising-Simons Foundation; the French Alternative Energies and Atomic Energy Commission (CEA); the Secretariat of Science, Humanities, Technology and Innovation (SECIHTI) of Mexico; the Ministry of Science, Innovation and Universities of Spain (MICIU/AEI/10.13039/501100011033), and by the DESI Member Institutions: \url{https://www.desi.lbl.gov/collaborating-institutions}. Any opinions, findings, and conclusions or recommendations expressed in this material are those of the author(s) and do not necessarily reflect the views of the U. S. National Science Foundation, the U. S. Department of Energy, or any of the listed funding agencies.

The authors are honored to be permitted to conduct scientific research on I'oligam Du'ag (Kitt Peak), a mountain with particular significance to the Tohono O’odham Nation.

\end{acknowledgments}

\vspace{5mm}
\facility{Mayall (DESI)}

\software{
  \texttt{iSEDfit} \citep{moustakas17a}, 
  \texttt{NumPy} \citep{harris20a.numpy},
  \texttt{SciPy} \citep{virtanen20a.scipy},
  \texttt{Numba} \citep{lam2015numba},
  \texttt{mpi4py} \citep{mpi4py},
  \texttt{cupy} \citep{cupy_learningsys2017},
  \texttt{astropy} \citep{astropy:2013, astropy:2018, astropy:2022},
  \texttt{Matplotlib} \citep{hunter07a},
  \texttt{seaborn} \citep{waskom21a}
}

\begin{appendix}

\section{Galaxy Basis Templates}
\label{appendix:galtemplates}

DESI observes galaxy targets spanning a broad range of redshifts and
intrinsic spectral properties, making it essential that our galaxy
templates represent this diversity (J. Moustakas et al. 2026, in
preparation). BGS \citep{BGS.TS.Hahn.2023} targets a magnitude-limited
sample at $0\lesssim z\lesssim0.6$ that encompasses the full range of
galaxy spectral types, from massive quiescent systems with old stellar
populations to actively star-forming galaxies with young stellar
populations and strong emission lines. LRG \citep{LRG.TS.Zhou.2023}
targets at $0.4\lesssim z\lesssim1.1$ are predominantly old,
metal-rich galaxies exhibiting a prominent $4000$-\AA\ break and
little ongoing star formation. And ELG \citep{ELG.TS.Raichoor.2023}
targets at $0.6\lesssim z\lesssim1.6$ are actively star-forming
galaxies characterized by strong, high-equivalent-width nebular
emission lines, particularly the \oiilam{} doublet.

To represent this diversity in galaxy type and astrophysical
properties, our strategy is to construct a large parent sample of
\textit{basis templates} grounded in observations of real galaxies.
The AGN and Galaxy Evolution Survey \citep[AGES;][]{Kochanek2012}
provides the BGS training sample at $0\lesssim z\lesssim0.75$; the
Deep Extragalactic Evolutionary Probe 2 \citep[DEEP2;][]{newman13c}
provides the ELG training sample at $0.75\lesssim z\lesssim1.45$; and
a photometric and spectroscopic compilation of color-selected galaxies
from the DESI Legacy Imaging Surveys Data Release 7
\citep[Legacy Surveys DR7;][]{Dey2019} provides the LRG training
sample at $0.2\lesssim
z\lesssim1.2$.\footnote{\url{https://www.legacysurvey.org/dr7}}

We generate the basis templates using a data-driven approach. We start
by modeling each object's available broadband photometry (at its
spectroscopic redshift) using stellar population synthesis (SPS)
models. SPS models yield a model of the stellar continuum at high
spectral resolution ($\mathcal{R}\approx10,000$), without the
instrumental effects of each original survey, and over a broad range
of rest-frame wavelength ($0.1$--$6~\micron$) extending well
beyond what survey spectroscopy alone provides. Then, we use empirical
correlations between observed nebular line ratios and the
$4000$-\AA\ break, $D_n(4000)$ (the ratio of the mean flux density
$F_\nu$ in the $4000$--$4100$~\AA\ and $3850$--$3950$~\AA\ rest-frame
windows), to tie this continuum model to an
inferred emission-line spectrum, yielding a semi-empirical description
of each galaxy's full spectral energy distribution. This
semi-empirical treatment of the emission lines avoids having to
predict line strengths from first-principles photoionization modeling,
which remains highly uncertain, particularly for forbidden-line ratios.

In \S\ref{appendix:sps} we describe this approach in more detail
as well as the SPS and emission-line modeling common to all three
template sets, and in \S\S \ref{appendix:bgs}, \ref{appendix:elg},
and \ref{appendix:lrg} we describe the specific construction of the
BGS, ELG, and LRG basis templates, respectively.

\subsection{Stellar Continuum and Emission-Line Modeling}
\label{appendix:sps}

The rest-frame UV/optical spectral energy distributions of galaxies
consist of two principal components: a stellar continuum arising from
the galaxy's composite stellar population, which encodes its star
formation and chemical evolution history, and a series of narrow
emission lines produced in \ion{H}{2} regions---$T\approx10^4$~K gas
photoionized by hot, young stars \citep{Kennicutt1992, conroy13a,
  kewley19a}. Galaxies with no ongoing star formation produce pure
stellar continua, while actively star-forming galaxies can exhibit
emission-line equivalent widths of hundreds to thousands of
Angstroms. In addition, dust grains attenuate both components by
wavelength-dependent amounts \citep[and references
  therein]{conroy13a}.

\begin{figure*}[t]
\centering
\includegraphics[width=\textwidth]{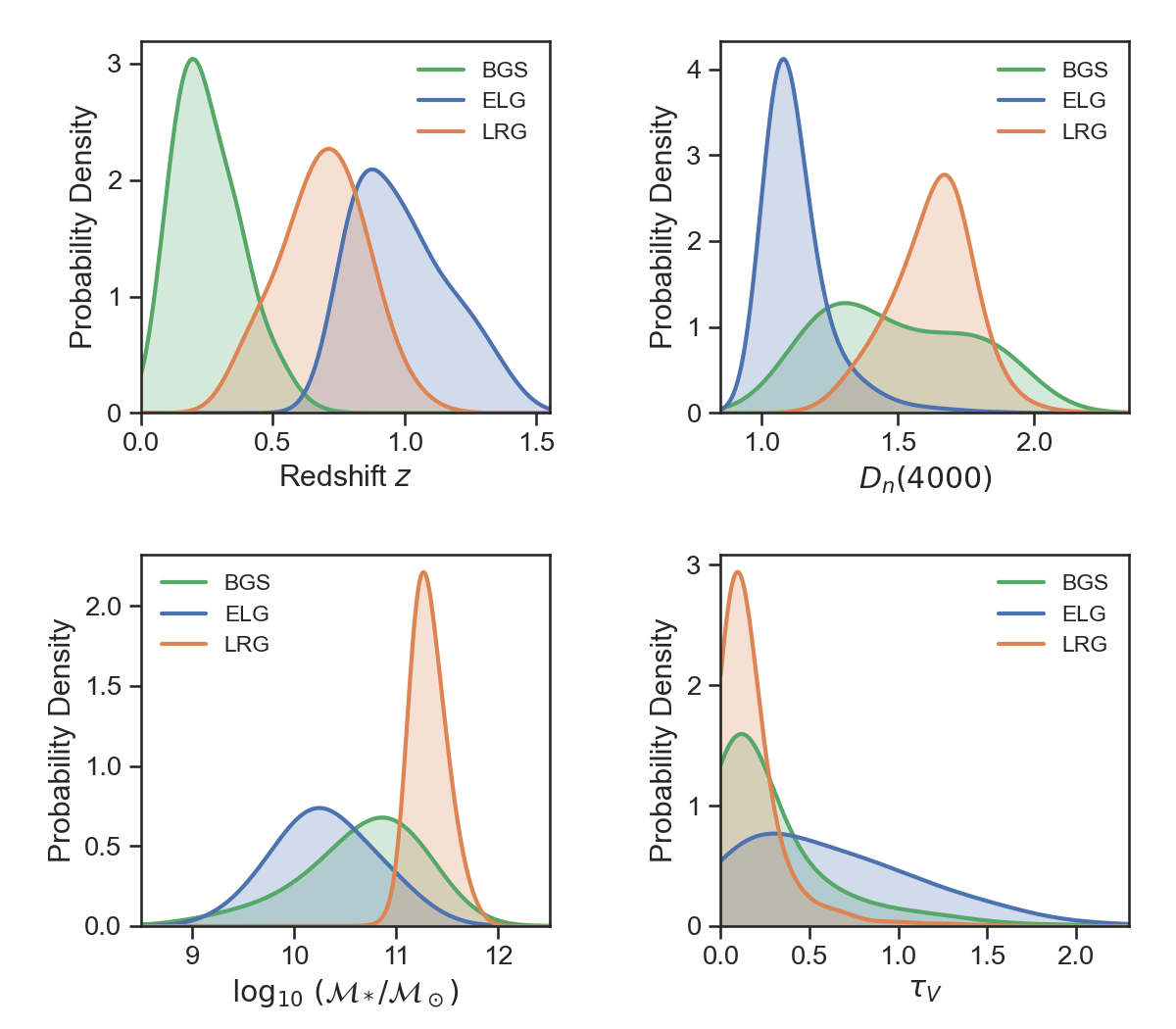}
\caption{Distributions of four physical properties for the BGS
  (green), ELG (blue), and LRG (orange) basis templates, derived from
  Bayesian SED modeling of the respective training samples (see
  \S\S \ref{appendix:bgs}--\ref{appendix:lrg}). The three classes
  of objects occupy complementary regions of parameter space: BGS
  spans the full range of stellar populations and dust content at low
  redshift; ELGs are young, low-mass, moderately dusty star-forming
  galaxies at intermediate redshift; and LRGs are massive, evolved,
  nearly dust-free galaxies at $z\approx0.2$--$1.1$.
  \label{fig:templateprops}}
\end{figure*}

For all three template sets, we model the stellar continuum of each
training galaxy using the \texttt{iSEDfit} Bayesian spectral energy
distribution (SED) fitting code \citep{moustakas13a,
  moustakas17a}.\footnote{\url{https://github.com/moustakas/isedfit}}
We build simple stellar populations (SSPs) from the theoretical
\texttt{C3\_K} stellar library (CKC14z; C. Conroy 2014, private
communication; see also \citealt{conroy18a}), \texttt{Padova} stellar
evolution isochrones \citep{girardi02a, marigo17a}, and a
\citet{kroupa01a} initial mass function (IMF) spanning
$0.1$--$100~\mathcal{M}_{\sun}$. We allow the stellar metallicity to
range from $Z=0.004$ to $0.03$ ($20\%$--$160$\% solar). The spectral
coverage of these models spans $0.1$--$1000~\micron$ at a resolution
of $\mathcal{R}=10,000$ between $0.15$ and $1~\micron$---more than
twice the maximum instrumental resolution of the DESI spectrographs
\citep{guy23a, abareshi22a}---ensuring that the models can be
convolved to match any instrumental line spread function without loss
of information. To model the dust attenuation, we use the simple
\citet{charlot00a} dust law,
$\tau(\lambda)\propto(\lambda/5500~\mathrm{\AA})^{-n}$, where
$\tau(\lambda)$ is the wavelength-dependent optical depth and we take
$n=0.7$.

For the BGS and ELG samples, we parameterize each galaxy's star
formation history (SFH) as an exponentially declining $\tau$-model,
$\psi\propto e^{-t/\tau}$, supplemented by stochastic, constant star
formation bursts of varying duration and burst fraction. We generate
30,000 Monte Carlo realizations of the model parameters spanning ages
$t\in[0.1, 12.5]$~Gyr, $e$-folding times $\tau\in[0,10]$~Gyr, and
assuming a $20\%$ burst probability. For the LRG training sample,
which predominantly consists of quiescent galaxies dominated by
evolved stellar populations, we adopt a delayed-$\tau$ SFH,
$\psi\propto te^{-t/\tau}$, $t\in[1.0, 11.05]$~Gyr, $\tau\in[0,
  6]$~Gyr, and no burst component. The maximum-likelihood SED model
for each galaxy yields its rest-frame stellar continuum spectrum at
$\mathcal{R}=10,000$, normalized to $1~\mathcal{M}_{\sun}$ at $10$~pc.

Figure~\ref{fig:templateprops} summarizes the distributions of four
physical properties---redshift, $D_n(4000)$, stellar mass, and dust
optical depth $\tau_V$---across all three basis template sets,
illustrating the complementary physical parameter space each class
occupies.

Next, we construct the integrated emission-line spectrum of each
object. Generating physically realistic emission-line spectra---across
the full rest-frame optical through near-infrared wavelength
range---is considerably more challenging to do from first principles
alone. The absolute and relative emission-line strengths depend on the
complex interplay between ionizing photon production, gas-phase
abundance, excitation conditions in the gas, and geometry, which are
difficult to infer even for individual \ion{H}{2} regions
\citep{stasinska03a, byler17a}. We therefore use an empirical approach
in which we measure the primary emission-line anchor for each template
class---\hbeta{} for BGS and \oii{} for ELGs---directly from the
observed spectrum of the training galaxy and then build the rest of
the emission-line spectrum using relations calibrated on local
star-forming galaxies and \ion{H}{2} regions.

\begin{figure}[t]
\centering
\includegraphics[scale=0.6]{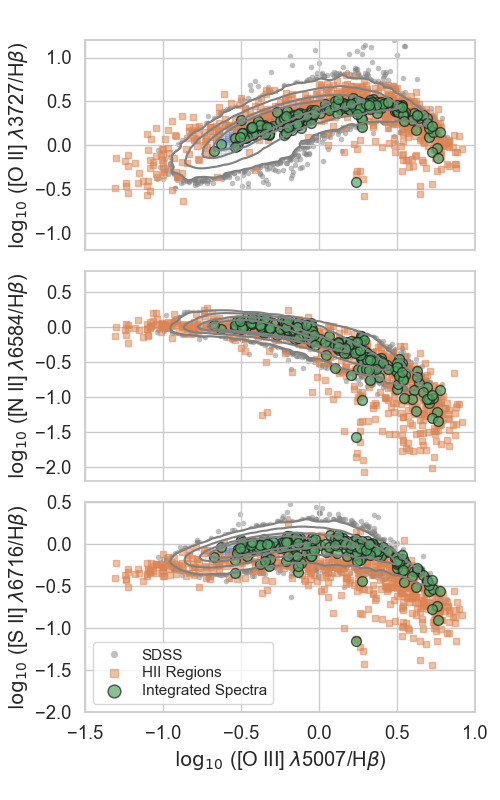}
\caption{Observed forbidden emission-line ratios relative to H$\beta$,
  used to calibrate the six-component GMM. Each panel shows a
  different line ratio versus
  $\log_{10}$([O~\textsc{iii}]~$\lambda5007$/H$\beta$): SDSS contours
  (gray; \citealt{brinchmann04a, tremonti04a}), extragalactic
  \ion{H}{2} regions (orange squares; \citealt{moustakas10a}), and
  integrated galaxy spectra of low-redshift star-forming galaxies
  (green circles; \citealt{moustakas06a, moustakas06b}). Only the
  \ion{H}{2} region and integrated spectral atlas data (807 objects
  total) enter the GMM fit; we show the SDSS contours for context
  only. The GMM assigns the fluxes of these four forbidden lines to
  the BGS and ELG basis templates (\S\S \ref{appendix:bgs}
  and \ref{appendix:elg}).
  \label{fig:lineratios}}
\end{figure}

The strongest forbidden lines in the optical are \oiilam{},
[N~\textsc{ii}]~$\lambda\lambda6548,84$, and
[S~\textsc{ii}]~$\lambda\lambda6716,31$, whose strengths relative to
H$\beta$ (and to one another) vary significantly with
[O~\textsc{iii}]~$\lambda\lambda4959,5007$/H$\beta$. Moreover, the
[O~\textsc{iii}] doublet is itself sensitive to gas-phase abundance
and excitation. To quantify these variations, we assemble a dataset of
619 extragalactic \ion{H}{2} regions \citep{moustakas10a}, 188
integrated galaxy spectra \citep[807 objects total;][]{moustakas06a,
  moustakas06b}, and $\approx10^{5}$ galaxies from the SDSS
\citep{brinchmann04a, tremonti04a}, and present the observed line-ratio
distributions in Figure~\ref{fig:lineratios}.

To model the four-dimensional distribution of
$\log_{10}$([O~\textsc{iii}]~$\lambda5007$/H$\beta$),
$\log_{10}$(\oii~$\lambda3727$/H$\beta$),
$\log_{10}$([N~\textsc{ii}]~$\lambda6584$/H$\beta$), and
$\log_{10}$([S~\textsc{ii}]~$\lambda6716$/H$\beta$), we use a
multi-component Gaussian mixture model (GMM) fitted to just the
\ion{H}{2} region and integrated galaxy spectra. We use the Bayesian
information criterion (BIC) to determine the number of GMM components
needed to explain the data without overfitting, eventually settling on
$N=6$ components.

With the GMM covariance matrices in hand, we proceed to construct each
galaxy's full emission-line spectrum using the following
procedure. First, we set the absolute strength of each galaxy's
``anchor line'' by multiplying the spectroscopically measured
rest-frame equivalent width by the continuum flux density at the
anchor wavelength predicted by that object's best-fitting SED
model. The anchor line for BGS and ELG targets is H$\beta$ and \oii,
respectively, both of which we measure for all AGES and DEEP2 spectra
used in our analysis (see \S\S \ref{appendix:bgs} and
\ref{appendix:elg}). This measured equivalent width also correlates
with the $4000$-\AA\ spectral break strength $D_n(4000)$, which traces
the age of the stellar population (Figure~\ref{fig:dn4000}); we
calibrate this relationship separately for each training sample in
\S\S \ref{appendix:bgs} and \ref{appendix:elg}, and use it in
\S\ref{appendix:galaxy-mc} to generate Monte Carlo realizations of the
basis templates.

Next, we use the empirically calibrated GMM to draw a random
realization of the four forbidden-to-H$\beta$ line ratios,
[O~\textsc{iii}]~$\lambda5007$, \oiilam{},
[N~\textsc{ii}]~$\lambda6584$, and
[S~\textsc{ii}]~$\lambda6716$. Because the GMM retains the full
covariance of the training data, each realization preserves the
physical correlations among the lines.

We then utilize the expected doublet ratio for each of these four
lines. Atomic transition probabilities fix the
[O~\textsc{iii}]~$\lambda5007/$[O~\textsc{iii}]~$\lambda4959$ and
[N~\textsc{ii}]~$\lambda6584/$[N~\textsc{ii}]~$\lambda6548$ ratios at
$2.8875$ and $2.93579$, respectively. The
[O~\textsc{ii}]~$\lambda3726/$[O~\textsc{ii}]~$\lambda3729$ and
[S~\textsc{ii}]~$\lambda6716/$[S~\textsc{ii}]~$\lambda6731$ doublets,
by contrast, are density-sensitive, but we fix them to a
representative value characteristic of the low-density limit---$0.73$
for \oii{} and $1.3$ for [S~\textsc{ii}]---appropriate for the
low-density \ion{H}{2}-region gas that dominates the nebular emission
in our training samples.

The templates also include three additional (albeit weak) forbidden
doublets---[O~\textsc{i}]~$\lambda\lambda6300,63$,
[S~\textsc{iii}]~$\lambda\lambda9069,9532$, and
[Ar~\textsc{iii}]~$\lambda\lambda7135,7751$, six lines in all---that
lie beyond the four line ratios modeled by the GMM; for simplicity, we
instead fix the overall strength of each doublet to a representative
ratio relative to H$\beta$ \citep{Kotulla2009}. As before, we assume,
for simplicity, fixed doublet ratios given by atomic transition
probabilities:
[O~\textsc{i}]~$\lambda6300/$[O~\textsc{i}]~$\lambda6363$ ($3.03502$),
[S~\textsc{iii}]~$\lambda9532/$[S~\textsc{iii}]~$\lambda9069$
($2.4686$), and
[Ar~\textsc{iii}]~$\lambda7135/$[Ar~\textsc{iii}]~$\lambda7751$
($4.16988$).


Finally, we compute the strength of the hydrogen and helium
recombination lines (H$\alpha$, H$\gamma$, H$\delta$, \ion{He}{1}, and
others; see Table~\ref{tab:emlines}), all scaled by the H$\beta$
line-strength, using standard Case~B recombination ratios assuming a
He/H abundance ratio of 0.0897 \citep{Osterbrock1989, Kotulla2009}.

\startlongtable
\begin{deluxetable}{ccc}
\tabletypesize{\footnotesize}
\tablecaption{Emission Lines\label{tab:emlines}}
\tablewidth{0pt}
\tablehead{
\colhead{Recombination} &
\colhead{Vacuum Wavelength} &
\colhead{Relative} \\
\colhead{Line} &
\colhead{(\AA)} &
\colhead{Flux}
}
\startdata
  Ly$\alpha$ & 1215.670 &    23.54656 \\
   H$\alpha$ & 6564.613 &    2.863158 \\
    H$\beta$ & 4862.683 &    1.000000 \\
   H$\gamma$ & 4341.684 &   0.4683401 \\
   H$\delta$ & 4102.892 &   0.2589474 \\
 H$\epsilon$ & 3971.195 &   0.1590283 \\
       H6 & 3890.151 &   0.1050202 \\
       H7 & 3836.472 &  0.07312550 \\
       H8 & 3798.976 &  0.05302830 \\
       H9 & 3771.701 &  0.03971660 \\
      H10 & 3751.217 &  0.03053440 \\
      H11 & 3735.430 &  0.02399190 \\
      H12 & 3722.997 &  0.01920650 \\
      H13 & 3713.027 &  0.01561130 \\
      H14 & 3704.906 &  0.01286640 \\
      H15 & 3698.204 &  0.01073680 \\
      H16 & 3692.605 & 0.009052600 \\
  Pa$\alpha$ & 18756.13 &   0.3386235 \\
   Pa$\beta$ & 12821.59 &   0.1632389 \\
  Pa$\gamma$ & 10941.09 &  0.09044530 \\
  Pa$\delta$ & 10052.13 &  0.05545750 \\
Pa$\epsilon$ & 9548.591 &  0.03656680 \\
      Pa9 & 9231.547 &  0.02543320 \\
     Pa10 & 9017.385 &  0.01842910 \\
     Pa11 & 8865.217 &  0.01378950 \\
      Pa12 & 8752.876 &  0.01059920 \\
  Br$\alpha$ & 40522.62 &  0.08021050 \\
   Br$\beta$ & 26258.67 &  0.04546560 \\
  Br$\gamma$ & 21661.20 &  0.02781380 \\
  Br$\delta$ & 19450.87 &  0.01825910 \\
{\ion{He}{1}}~$\lambda3188$ & 3188.662 &  0.04076080  \\
{\ion{He}{1}}~$\lambda3889$ & 3889.752 &   0.1027011  \\
{\ion{He}{1}}~$\lambda3965$ & 3965.852 &  0.01022650  \\
{\ion{He}{1}}~$\lambda4026$ & 4027.348 &  0.02115760  \\
{\ion{He}{1}}~$\lambda4471$ & 4472.755 &  0.04445050  \\
{\ion{He}{1}}~$\lambda4922$ & 4923.304 &  0.01199870  \\
{\ion{He}{1}}~$\lambda5016$ & 5017.079 &  0.02568980  \\
{\ion{He}{1}}~$\lambda5876$ & 5877.299 &   0.1186800  \\
{\ion{He}{1}}~$\lambda6678$ & 6679.994 &  0.03360660  \\
{\ion{He}{1}}~$\lambda7065$ & 7067.198 &  0.02173860  \\
{\ion{He}{1}}~$\lambda10830$ & 10833.22 &   0.2404832  \\
\enddata
\tablecomments{Rest-frame vacuum wavelengths and flux ratios for all hydrogen
  and helium recombination lines included in the basis templates relative to H$\beta=1.0$.
}
\end{deluxetable}

We model each of these emission lines as a single Gaussian
centered on its rest-frame vacuum wavelength, so the lines share the
systemic redshift of the stellar continuum, with no relative velocity
offsets, asymmetries, or broad components. All the lines in a given
spectrum have the same intrinsic velocity width
(\S\ref{appendix:galaxy-mc}).

The following three sections apply this shared SPS and emission-line
framework to each target class in turn, describing the training
sample, selection cuts, and template-specific calibrations we use to
construct the BGS, ELG, and LRG basis templates.

\subsection{BGS Basis Templates}
\label{appendix:bgs}

We construct the BGS basis templates from AGES, which obtained
intermediate-resolution optical spectroscopy of an $I_{\rm AB}<20.45$
magnitude-limited sample of $\approx11,000$ galaxies at $0\lesssim
z\lesssim0.8$ over $\approx8$~deg$^2$ using the Hectospec multi-fiber
spectrograph on the MMT. The AGES spectra span approximately
$3700$--$9200$~\AA\ at $\mathcal{R}\approx1000$
($1.2$~\AA\ pixel$^{-1}$), providing access to H$\beta$ throughout the
full survey redshift range. Each galaxy has aperture-matched broadband
$UB_{\rm W}RIzJHK_s$ photometry, unWISE $W1$ and $W2$
forced-photometry at $3.4$ and $4.6~\micron$
\citep{Lang2016}\footnote{\url{https://unwise.me}}, and a statistical
weight correcting for photometric, targeting, and redshift
incompleteness. We measure emission-line fluxes and equivalent widths
for the full AGES sample, including H$\beta$, H$\alpha$, and the
principal forbidden lines, using the spectral-fitting methods
described in \citet{moustakas10a, moustakas11a}.

From the full AGES catalog, we select the BGS training sample by applying
the following cuts:

\begin{itemize}
  \item $0.05 < z < 0.75$;
  \item $I_{\rm NDWFS} < 20.4$ (AB, extinction-corrected); and
  \item photometric detections in at least six broadband filters.
\end{itemize}

\begin{figure}[t]
\centering
\includegraphics[scale=0.6]{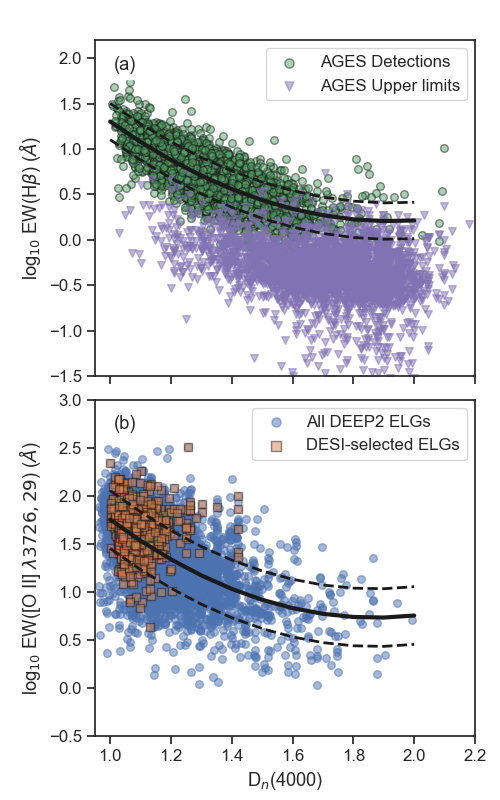}
\caption{Rest-frame emission-line equivalent width (EW) versus
  $D_n(4000)$ break strength for the BGS and ELG training samples.
  \textit{(a)}: EW(\hbeta) for the BGS training sample from AGES;
  green circles are \hbeta-detected objects and purple triangles are
  $1\sigma$ upper limits. \textit{(b)}: EW(\oii~$\lambda\lambda3726,29$)
  for the ELG training sample from DEEP2; blue circles are the full
  sample and orange squares are the subset satisfying DESI ELG
  target-selection color cuts \citep{ELG.TS.Raichoor.2023}. In each panel the
  solid curve shows the best-fit second-order polynomial
  (Equations~\ref{eq:bgs_ew} and~\ref{eq:elg_ew}), and dashed curves
  show the $\pm1\sigma$ intrinsic scatter.
  \label{fig:dn4000}}
\end{figure}

\noindent In Figure~\ref{fig:dn4000}a we show the correlation between
EW(\hbeta) and $D_n(4000)$ for this sample of $7636$ galaxies. The
filled green circles show the $4187$ objects ($55\%$) where \hbeta{}
is detected with ${\rm S/N} > 3$ and rest-frame ${\rm EW}({\rm
  H}\beta) > 1$~\AA, while the open purple triangles show the
remaining $3449$ galaxies ($45\%$) where \hbeta{} is an upper
limit. For the subset of objects where \hbeta{} is well-detected, we
fit the correlation with a second-order polynomial,
\begin{equation}
\log_{10}\,{\rm EW}({\rm H}\beta) = \sum_{i=0}^{2} a_{i}\, D_{n}(4000)^{i},
\label{eq:bgs_ew}
\end{equation}

\noindent where $\mathbf{a}=\{1.28520974, -4.94408026, 4.9617704\}$
with a scatter of approximately $0.2$~dex. In
\S\ref{appendix:galaxy-mc} we describe how we use this fit to
generate Monte Carlo template realizations of BGS-like templates.

Finally, we use the methodology described in
\S\ref{appendix:sps} to construct a complete high-resolution SED
of each object, using \hbeta{} (where it is well-detected) as the
anchor line we use to generate the emission-line spectrum. The
resulting $7636$ BGS basis templates span $0.1$--$6~\micron$ at
$\mathcal{R}\approx10,000$ and a pixel size of $50$~km~s$^{-1}$. These
templates are publicly released as \texttt{bgs\_templates\_v2.3.fits}.

\subsection{ELG Basis Templates}
\label{appendix:elg}

We construct the ELG basis templates from DEEP2/DR4 \citep{newman13c},
which obtained high-resolution optical spectroscopy of more than
50,000 galaxies using the DEIMOS multi-object spectrograph on the
Keck~2 telescope. The spectra span approximately
$6500$--$9300$~\AA\ at $\mathcal{R}\approx6000$
($0.33$~\AA\ pixel$^{-1}$), fully resolving the \oiilam\ doublet for
galaxies at $z\approx0.7$--$1.4$. From the public DR4 catalog, we
select 35,682 galaxies with high-quality redshifts ($Q\geq3$;
\citealt{newman13c}) and well-calibrated one-dimensional spectra. We
assemble a photometric catalog for each galaxy by matching to the
extended photometric catalog of \citet{matthews13}, which provides
$BRI$ photometry and $ugriz$ photometry from the CFHTLS surveys (DEEP2
Field~1) or SDSS (Fields~2--4), and add WISE $W1$ and $W2$ photometry
from unWISE forced-photometry measurements at the positions of DEEP2
sources brighter than $R=24.1$ \citep{Lang2016}.

We model the broadband photometry with \texttt{iSEDfit}
(\S\ref{appendix:sps}), fitting Field~1 and Fields~2--4 with
appropriate filter transmission curves but otherwise identical SPS
modeling priors. We measure emission-line fluxes and equivalent widths
using the spectral-fitting methods of \citet{moustakas10a,
  moustakas11a}, fixing the \oii{} $\lambda3726/\lambda3729$ flux
ratio to its measured median value of $0.73$. We consider an emission
line detected if its amplitude exceeds $1.5$ times the local continuum
rms; otherwise we record a $1\sigma$ upper limit. However, because the
DEEP2 spectra are only approximately flux-calibrated, we derive the
integrated \oii{} flux for each galaxy following \citet{zhu09a},
multiplying the rest-frame equivalent width from the spectral fit by
the continuum flux density at $3727$~\AA{} estimated from the
best-fitting \texttt{iSEDfit} SED model.

From the full DEEP2 catalog, we select the ELG training sample by
applying the following cuts:
\begin{itemize}
  \item $0.75 < z < 1.45$;
  \item $r_{\rm CFHTLS} < 23.5$ (AB, extinction-corrected);
  \item SED-fitting $\chi^2 < 40$ with photometric detections in at least six broadband filters;
  \item \oii\ doublet detected with ${\rm S/N} > 3$; and
  \item EW(\oiilam)$> 1$~\AA.
\end{itemize}

\noindent These cuts result in a sample of $7,735$ galaxies. In
Figure~\ref{fig:dn4000}b we show the relationship between EW(\oii) and
$D_n(4000)$ for this sample (filled blue circles) and, for reference,
the subset of objects which satisfy the DESI ELG color-cuts
\citep{ELG.TS.Raichoor.2023}. As for the BGS training sample, we fit the
relationship with a second-order polynomial,
\begin{equation}
\log_{10}\,{\rm EW}(\text{\oii}~\lambda\lambda3726,29) = \sum_{i=0}^{2} b_{i}\, D_{n}(4000)^{i},
\label{eq:elg_ew}
\end{equation}
where $\mathbf{b}=\{1.34323087, -5.02866474, 5.43842874\}$, with an
intrinsic scatter of approximately
$0.3$~dex. \S\ref{appendix:galaxy-mc} describes how we use this
fit to generate Monte Carlo template realizations.

Finally, we generate the full integrated stellar continuum and
emission-line spectrum for every galaxy using the methodology
described in \S\ref{appendix:sps} with \oii{} as our anchor
emission line. The resulting $7,735$ ELG basis templates share the
same wavelength grid as the BGS templates: $0.05$--$6~\micron$ at
$\mathcal{R}\approx10,000$ and $50$~km~s$^{-1}$ pixels. These templates
are publicly released as \texttt{elg\_templates\_v2.2.fits}.

\subsection{LRG Basis Templates}
\label{appendix:lrg}

We construct the LRG basis templates from a spectroscopic compilation
of galaxies assembled to be statistically representative of the
photometric LRG population targeted by DESI
\citep{LRG.TS.Zhou.2023}. We begin by selecting 111,114 DESI-like LRGs
at $0.2\lesssim z\lesssim1.2$ using the Legacy Surveys DR7 $grzW1W2$
photometry and the \citet{Zhou2021} photometric redshift catalog. We
cross-match this sample against a wide range of public spectroscopic
catalogs from SDSS, (e)BOSS, AGES, DEEP2, GAMA, OzDES, VIPERS, VVDS,
and COSMOS to identify 75,315 LRGs satisfying the following
requirements:
\begin{itemize}
  \item $0.2 < z < 1.2$;
  \item at least three observations in each of the $g$, $r$, and $z$ bands
    (${\tt nobs}_{grz}\geq3$);
  \item measured flux in all five $grzW1W2$ bands.
  \item not flagged as a WISE artifact or lying near a bright star
    (${\tt wisemask}={\tt brightstarinblob}={\rm False}$); and
\end{itemize}

Because this spectroscopic sample is not uniform across the sky, we
assign each object a statistical weight proportional to the number of
photometric LRG candidates within its neighborhood in the joint space
of redshift, apparent magnitude, and observed color. This weight
ensures that the final template set statistically mirrors the
photometric parent population of $>3.3$~million LRGs with precise
photometric redshifts \citep{Zhou2021}.

We model the $grzW1W2$ photometry at the spectroscopic redshift with
\texttt{iSEDfit} (\S\ref{appendix:sps}), adopting the
delayed-$\tau$ star-formation history appropriate for this
predominantly quiescent population. We do not include emission lines
in this sample, even though a small fraction of LRGs are expected to
exhibit weak line-emission typically due to residual star formation
\citep[e.g.,][]{zhu10a}. From the fitted sample, we select 3,000 basis
templates via a probability-weighted random draw, in which each galaxy
is selected with probability proportional to its statistical weight;
this method preferentially samples galaxies that best represent the
photometric LRG population while maintaining diversity across the full
parameter space of the spectroscopic training set.

The resulting $3,000$ LRG basis templates share the same wavelength
grid as the BGS and ELG templates: $0.05$--$6~\micron$ at
$\mathcal{R}\approx10,000$ and $50$~km~s$^{-1}$ pixels. These templates
are publicly released as \texttt{lrg\_templates\_v2.2.fits}.

\subsection{Monte Carlo Template Realizations}
\label{appendix:galaxy-mc}

The basis templates described in this appendix represent a fixed,
discrete set of real training galaxies. To build the larger,
continuous ensemble of synthetic spectra needed to derive the final
galaxy PCA templates (\S\ref{sec:galaxy-templates})---10,000
ELG, 5,000 LRG, and 5,000 BGS realizations---we use the \texttt{BGS},
\texttt{ELG}, and \texttt{LRG} classes of the \texttt{desisim}
software package,\footnote{\url{https://github.com/desihub/desisim}}
each of which subclasses a common \texttt{GALAXY} base class and
exposes a \texttt{make\_templates()} method for generating any number
of Monte Carlo realizations from a single input random seed.

For each realization, \texttt{make\_templates()} draws a random
redshift and apparent magnitude from class-specific ranges, then
randomly selects one of the basis-template continua from
\S\S \ref{appendix:bgs}--\ref{appendix:lrg}. For BGS and ELG
targets, it also draws a new synthetic anchor-line equivalent width
from the calibrated $D_n(4000)$ relation (Figure~\ref{fig:dn4000}) and
its intrinsic scatter, together with a new realization of the four
forbidden-to-H$\beta$ line ratios from the GMM described in
\S\ref{appendix:sps}. Both draws are independent of the values
used for the parent basis template, decoupling the number of Monte
Carlo emission-line realizations from the fixed number of underlying
continua. For a BGS realization built from an H$\beta$ upper-limit
basis template (\S\ref{appendix:bgs}), this synthetic equivalent
width is instead set to zero, so the realization retains the parent
galaxy's stellar continuum without injected emission lines. LRG
templates skip the emission-line draw entirely, since the LRG basis
templates contain no emission lines. Each realization also draws a
velocity dispersion uniformly from 50--150~km~s$^{-1}$ (ELG),
120--300~km~s$^{-1}$ (BGS), or 150--300~km~s$^{-1}$ (LRG), which sets
the width of every emission line and broadens the stellar
continuum. \texttt{make\_templates()} then
assembles the resulting spectrum, optionally applies the target
class's DESI selection color-cuts, and repeats the process until it
produces the requested number of passing models. A single input seed
fully determines every random draw, making the realizations fully
reproducible.

\section{Intrinsic velocity shifts of quasar emission lines}
\label{appendix:qso_lineshifts}

The non-uniform velocity shifts of quasar emission lines relative to their host galaxy are a notorious limitation 
to redshift accuracy. Owing to the finite number of principal components, a set of PCA quasar templates cannot fully 
capture the diversity of outflow properties (and thus emission line profiles) present in the DESI spectroscopic
sample. Further, because of their empirical nature, the templates will incorporate any redshift biases present in the training
sample. 

The quasar templates used in this manuscript were derived from the SDSS Data Release 16 quasar catalog \citep[DR16Q;][]{Lyke2020-DR16Q}. The 
redshift biases of this sample were mitigated by correcting individual spectra to the measured Mg~\textsc{ii} restframe 
\citep{Brodzeller2023.QSOTemplates}, chosen for its relative stability compared with other quasar emission lines 
\citep[e.g.,][]{shen16,hewett2010} and visibility over a wide redshift range. This line, however, is not a perfect indicator of systemic redshift, often exhibiting a velocity shift on the order of tens of km~s$^{-1}$. This appendix explores the accuracy of quasar redshift estimates obtained with these 
templates for DESI through a spectral stacking analysis.

We use spectral stacking to enhance the signal-to-noise of weak features associated with quasar host galaxies, such as
[O~\textsc{ii}] and Ca~\textsc{ii}~K. The offset of these features relative to the expectation in the absence of redshift 
error estimates the average redshift accuracy for the stacked sample. We also measure the offset of the quasar emission lines Mg~\textsc{ii} and C~\textsc{iv}, since at $z\gtrsim1.5$ only broad lines are observable, thus these drive the redshift estimates. The offsets for these two features cannot be used directly as estimators of accuracy, however, because they incorporate an intrinsic velocity shift from the systemic redshift. Instead, we use Mg~\textsc{ii}'s average offset from Ca~\textsc{ii}~K as an estimate for the average outflow velocity for Mg~\textsc{ii}, and use that to extrapolate the average redshift accuracy at higher redshifts when galaxy features are no longer visible. We include C~\textsc{iv} only to illustrate the complex quasar dynamics that challenge template redshift estimation, particularly at high redshifts.

We use the DR2 Ly$\alpha$ quasar catalog \citep[][]{DESI2025.Lya}, which incorporates information from the quasar 
afterburner algorithms to reduce catastrophic failures caused by line confusion \citep[see][]{QSO.TS.Chaussidon.2023}. 
Although additional algorithms are used to construct this catalog, all final redshifts come from Redrock. We exclude known
BAL quasars \citep{filbert24}, since this sub-class introduces additional redshift uncertainties unrelated to the PCA 
template bias and outflow velocities within the sample \citep[e.g.,][]{garcia23}. 

 We bin the quasar catalog on $\Delta z = 0.1$ over $0.35<z<3.5$, and compute the error-weighted mean spectrum of each bin after shifting the individual spectra to the Redrock rest-frame solution. We fit the
 mean spectra in the regions of the Ca~\textsc{ii}~K, [O~\textsc{ii}], Mg~\textsc{ii}, and C~\textsc{iv} emission lines (absorption line in the case of Ca~\textsc{ii}~K), if visible for
 the given redshift bin. The fit to each line consists of a power-law to describe the underlying continuum plus a multi-component Gaussian for the line. The fits are visually inspected to confirm quality, and the number of Gaussians is fixed to one 
 for Ca~\textsc{ii}~K and [O~\textsc{ii}], and two for Mg~\textsc{ii} and C~\textsc{iv}.

\begin{figure}[t]
\centering
\includegraphics[width=\columnwidth]{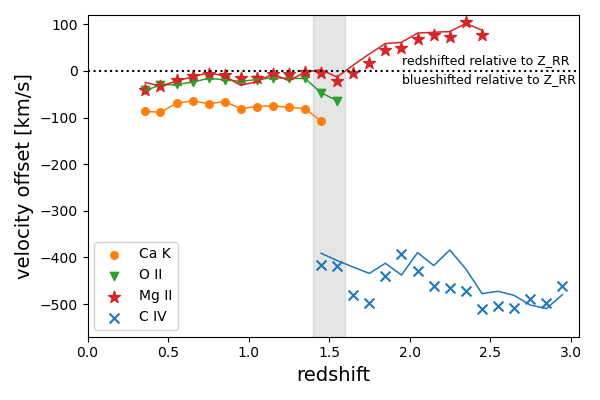}
\caption{Average velocity offset of prominent emission lines relative to the Redrock redshift estimates in various redshift bins as measured by the line centroid (markers) and the location of the line's peak flux (line). The grey-shaded region indicates the overlap between the LOZ and HIZ quasar templates, where either template set can provide the best-fit redshift solution. }
\label{fig:lineshifts}
\end{figure}

We measure the average velocity offset of each line from its expected rest-frame wavelength in the absence of redshift errors --- and absence of intrinsic shifts in the case of broad lines --- in two ways. First, using the difference between the centroid of the narrow Gaussian and the vacuum wavelength (Table~\ref{tab:qso_lines}), and then using the location of the peak flux of the emission line model described in the previous paragraph. The results are shown in Figure~\ref{fig:lineshifts}. The two methods yield consistent results for the galaxy features and are within $\sim$15 km~s$^{-1}$ for Mg~\textsc{ii}.

\begin{table}
\begin{tabular}{l|l}
Line & Vacuum Wavelength [\AA]\\
\hline
Ca~\textsc{ii}~K    &  3934.78 \\
$[$O~\textsc{ii}$]$    &  3728.48  \\
Mg~\textsc{ii}   &  2798.75  \\
C~\textsc{iv}     &  1549.06  \\
\end{tabular}
\caption{The vacuum rest-frame wavelength used for calculating line offsets in Appendix~\ref{appendix:qso_lineshifts}, taken from \cite{vandenberk01}.}
\label{tab:qso_lines}
\end{table} 

For the following discussion, we assume Ca~\textsc{ii}~K to be a reliable indicator of true redshift. That is, the velocity offset measured for Ca~\textsc{ii}~K can be entirely attributed to redshift errors --- a reasonable assumption given that it originates from stellar absorption.
With this in mind, the LOZ quasar templates ($0<z<1.6$) display a consistent bias of roughly $-75$~km~s$^{-1}$. In other words, redshifts are overestimated by this magnitude on average with the LOZ quasar templates. This is unsurprising, given that Mg~\textsc{ii} was the preferred rest-frame for the templates' parent sample. The preferred rest frame is further illustrated by the near-zero velocity offset of Mg~\textsc{ii} across the full LOZ redshift range. Informed by this study, a new version of the LOZ templates (v1.1) was produced for future DESI data releases, in which a positive velocity shift was applied to the wavelength grid, making Ca~\textsc{ii}~K more consistent with a zero velocity offset.

The transition from LOZ to HIZ is apparent in Figure~\ref{fig:lineshifts} with a clear shift in bias level. At $z\gtrsim1.5$, prominent emission features of the host galaxy are redshifted beyond the wavelength coverage of the DESI spectrograph. Thus, Redrock must rely on sufficient flexibility in the quasar templates to accommodate emission-line diversity while anchoring to the systemic redshift. If the difference between the velocity offset of Mg~\textsc{ii} and that of Ca~\textsc{ii}~K is roughly constant for the sample ($\sim$75~km~s$^{-1}$), this implies excellent average accuracy for HIZ redshifts. We note that there is still relatively large scatter in redshift error (Figure~\ref{fig:redshift_precision}), but the average redshift inaccuracy is expected to be rather small. This is supported by the analysis from \cite{Bault24a.QSOz} who reported an average redshift inaccuracy for DESI quasars consistent with zero at $-8\pm42$~km~s$^{-1}$ at $z>2$ with these HIZ templates through cross-correlation with the Ly$\alpha$ forest (their Figure~9).

\end{appendix}

\bibliographystyle{aasjournalv71}
\bibliography{biblio}

\end{document}